# On the nascent wind of nearby oxygen-rich AGB stars: a brief review

P. Darriulat, D.T. Hoai, P.T. Nhung, P.N. Diep, N.B. Ngoc, T.T. Thai, P. Tuan-Anh

*Department of Astrophysics, Vietnam National Space Center, Vietnam Academy of Science and Technology,*
*18, Hoang Quoc Viet, Nghia Do, Cau Giay, Ha Noi, Vietnam.*
*Email: darriulat@vnsc.org.vn (PD), dthoai@vnsc.org.vn (DTH), pttnhung@vnsc.org.vn (PTN)*

**Abstract**

The commonly accepted mechanism governing the formation of the nascent wind in oxygen-rich AGB stars combines an initial boost above the photosphere, given by shock waves resulting from stellar pulsations and convective cell granulation, with a subsequent acceleration fuelled by the radiation pressure of the star on dust grains. We use six nearby stars, for which detailed studies of visible and infrared observations at the VLT and millimetre observations at ALMA are available, to assess the extent to which the validity of this picture is currently corroborated. We show that while providing a very useful guide to current research and having received general support and suffered no contradiction, it still requires many additional observations to be reliably validated. In particular, observations of the highest possible angular resolution at both millimetre and visible/infrared wavelengths, performed in conjunction with measurements of the light curve, are necessary to tell apart the respective roles played by convection and stellar pulsations. The observed concurrence of high variability near the photosphere with persistence over decades, or even centuries, of the global anisotropy displayed by the CSE need to be understood. New observations of the close neighbourhood of the star are required to elucidate the mechanism that governs rotation, in particular in the cases of R Dor, $L_2$ Pup and EP Aqr. We argue that the presence of stellar or planetary companions does not seriously impact the formation of the nascent wind and only modifies its subsequent evolution.

# 1. Introduction
## 1.1 State-of-the-art

Much progress has been made over the past decade in understanding the mechanisms at stake in the formation of the nascent wind of oxygen-rich AGB stars. The most salient features have recently been described by S. Höfner, H. Olofsson and B. Freytag in a series of recent papers, in terms that we summarize below; we refer the reader to these articles for a list of relevant references, which can easily be traced from Freytag, Liljegren & Höfner (2017), Höfner & Olofsson (2018), Höfner & Freytag (2019) and Freytag and Höfner (2023). For simplicity, in the remainder of the article, we refer to their work as "the Standard Model".

In contrast to carbon-rich, oxygen-rich AGB stars have a dusty envelope that is more transparent in the visible and near-IR, giving a better view of the innermost dust-forming atmospheric layers. Their study supports a scenario where the radiation pressure, which triggers the outflows, is caused by photon scattering on highly transparent iron-free dust grains, which condense at shorter distances than iron-bearing silicate grains. As long as they are of a size comparable to the radiation wavelength, ~0.1-1.0 µm, they uphold sufficient radiation pressure. Studies of dust formation suggest (e.g. Decin et al. 2017, 2018, Takigawa et al. 2017, 2019, Ohnaka et al. 2016) that the condensation of corundum ($Al_2O_3$), forming a thin gravitationally bound dust layer close to the stellar photosphere, precedes the formation of silicate dust, the condensation of silicate mantles on top of corundum cores speeding up grain growth to the critical size.

Recent high-resolution imaging of nearby AGB stars at visible and infrared wavelengths has revealed complex, non-spherical distributions of gas and dust in the close circumstellar environment, with changes in morphology and grain sizes occurring over the course of weeks or months. Such inhomogeneous distribution of atmospheric gas emerges naturally in 3-D hydrodynamical models, as a consequence of large-scale convective flows below the photosphere and the resulting network of atmospheric shock waves, the dynamical patterns in the gas being imprinted on the dust in the close stellar environment. Global AGB star models (Liljegren et al. 2018) are characterized by giant convection cells, which can span over a steradian, and have a lifetime of many years. The photospheric layers have a complex, variable appearance due to smaller, shorter-lived cells, which form close to the surface of the convection zone. In addition to the convective flows, radial pulsations occur with typical periods at year scale. These dynamical processes generate waves of various frequencies and spatial scales, which quickly develop into shock waves as they propagate outward through the atmosphere with its steeply declining density. The shocks give rise to ballistic gas motions, which typically peak around 2 stellar radii. As the shock waves interact and merge, they produce large-scale regions of enhanced densities in their wakes.

As remarked by Höfner & Freytag (2019), the basic cloud-formation mechanism precedes wind acceleration. Its study is therefore a necessary preliminary to the understanding of the mass-loss mechanism; in this context, the acceleration of the wind, its possible subsequent interaction with planetary or stellar companions, and more generally the evolution toward the planetary nebula phase, appear as complications that can be ignored. This early phase of formation of the nascent wind appears instead as fundamental and its study deserves being dedicated major effort on both theoretical and observational fronts. Recently, two main approaches have been pursued toward this aim: observations in the visible and infrared, essentially using the VLT, and millimetre observations, essentially using ALMA.

The former, using the VLT in single-dish or interferometer mode, take advantage of the high performance of instruments such as SPHERE-ZIMPOL. Particularly remarkable are observations of π1 Gruis by Paladini et al. (2018) using PIONIER, which show evidence for the presence of large granulation cells on the stellar surface. In this context, using the same instrument, one must recall the observation of granulation on the surface of Betelgeuse (Montargès et al. 2016) and its confirmation from spectro-polarimeter observations (López Ariste et al. 2018) using NARVAL at Pic du Midi; the latter give evidence for slow evolution of the granule pattern on a 2 to 3 years timescale, accompanied by a much faster variability with small amplitude variations of a week to month time scale. NARVAL observations have also given evidence for asymmetric shocks in χ Cygni, induced by the interplay between pulsation and convection, suggesting the presence of a highly anisotropic and variable velocity field near the surface of the star (López Ariste et al. 2019).

In general, evidence for clumpy dust clouds near the star has been obtained from the detection of scattered light and multi-epoch observations have shown changes in morphology on month time scales. They have provided evaluations of the dust grain size consistent with the picture described above.

Millimetre observations have concentrated on three fronts: 1) study of the continuum emission of the stellar disc and surrounding hot dust, giving occasional evidence for hot spots and variability; 2) measurements of the radial dependence of the abundance of molecular species giving evidence, in particular, for the early formation of aluminium-rich dust grains; 3) measurements of the morpho-kinematics of the inner CSE layer using the emission of molecular lines.

*1.2 Six selected stars*

We have selected a sample of six representative oxygen-rich AGB stars to illustrate the most salient features displayed by the morpho-kinematics of their inner CSE layers. All stars in the sample are oxygen-rich nearby stars, with spectral type M and within 100 pc or so from the Sun. Their masses cover between ~0.7 and ~2 solar masses. Most are long period variables with periods at year scale (Pojmanski et al. 2005), with the exception of EP Aqr and $L_2$ Pup, which have periods of only 110 and 141 days, respectively. The period-luminosity relation shows that EQ Aqr pulsates on the first overtone while all others pulsate in the fundamental mode. They are shared between Mira and semi-regular types, but the separation between the two classes is sometimes unclear (e.g. W Hya). None shows a clear technetium signal in its spectrum (Little et al. 1987) but *o* Cet (Vanture et al. 1991, Kipper1992) and $L_2$ Pup (Lebzelter & Hron 1999) may display some. All have low mass-loss rates, not exceeding ~$2\times10^{-7}$ solar masses per year. *o* Cet has a clearly identified companion of mass 0.7 solar masses with a separation of ~80 au and displays features characteristic of symbiotic systems. Kervella et al. (2016) have claimed the presence of a planetary companion distant by only 2 au from the centre of $L_2$ Pup but this needs to be confirmed. Important parameters are listed in Table A1 of the Appendix, together with references of relevance to the present article. Table A2 lists molecular lines considered in the present review and observed with high angular resolution, together with important parameters and associated beam sizes.

At first glance, each of the oxygen-rich AGB stars that have been studied in some detail displays properties of its own making it appear as somewhat singular: unveiling common properties that can be considered as characteristic of the whole family is therefore very challenging. The aim of the present article is to make a step in this direction.

## 2. The stellar disc and its immediate environment

ALMA observations of continuum emission have provided images of the stellar disc and of its immediate environment at millimetre wavelengths, often displaying variability and hot spots interpreted as the effect of shocks induced by pulsations and convective cell granulation.

The disc of *o* Cet has been shown to have a 10% to 20% ellipticity in early measurements (Vlemmings et al. 2015, Matthews et al. 2015 at JVLA, see also Chandler et al. 2007 at ISI and Perrin et al. 2020 at IOTA) but was found circular later on (Wong et al. 2016, Vlemmings et al. 2019a). Kamínski et al. (2016), also using APEX and Herschel observations, have given evidence for strong inhomogeneity (hot spot) and irregular variability in the emission of AlO molecules. Spectro-polarimeter observations using NARVAL at Pic du Midi (Fabas et al. 2011) have given evidence for shock-induced polarized hydrogen emission lines, suggesting their association with convective cells.

Strong hot spots have been observed on W Hya, with lifetimes at the scale of weeks (Vlemmings et al. 2017 & 2019a) and on R Dor (Vlemmings et al. 2019a), correlated with dust structures observed in the visible. However, no hot spot has been directly observed on R Leo and EP Aqr; in the former case, Vlemmings et al. (2019a) have given evidence for a radial asymmetric expansion with mean velocity of 10.6 ±1.4 km s$^{-1}$ within 1-2 stellar radii; in the latter case, Homan et al. (2020a) have given evidence for a 20% flux density increase between 2016 and 2019, suggesting variability and sporadic contributions on scales larger than the stellar disc.

Observations using the VLT, at optical and infrared wavelengths, have given evidence for a dusty layer surrounding the photosphere, commonly displaying both inhomogeneity and variability. In particular SPHERE/ZIMPOL polarization measurements are available for $L_2$ Pup (Kervella et al. 2015), for *o* Cet (Khouri et al. 2018), for W Hya (Ohnaka et al. 2016 & 2017, Khouri et al. 2020), for R Dor (Khouri et al. 2016a) and for EP Aqr (Homan et al. 2020a). The degree of linear polarization usually reaches high values, well in excess of 10%, except for *o* Cet, where it does not exceed 4%, dust being confined at the border of a dense gas reservoir, within 100-200 mas from the centre of the star. In the case of W Hya, together with observations using NACO (Norris et al. 2012), AMBER/VLTI (Ohnaka et al. 2016,2017; Hadjara et al.

2019) and MIDI/VLTI (Zhao-Geisler et al. 2015), evidence is obtained for a clumpy and dusty layer, displaying important variability at short time scale. In the case of R Dor, Khouri et al. (2016a) find evidence for variability and asymmetry, with outflowing dust of steeply decreasing radial density and Ohnaka et al. (2019), using VLTI/AMBER, find evidence for sudden outward acceleration to 7-15 km s$^{-1}$ at radial distances between 1.5 and 1.8 stellar radii. In the case of EP Aqr (Homan et al. 2020a), the linearly polarised flux map suggests the presence of a torus of dust with an axis close to the known axi-symmetry axis of the CSE. Finally, the close neighbourhood of R Leo has been observed in the near-infrared by Fedele et al. (2005) using VLTI/VINCI, giving evidence for significant deviation from a uniform disc profile, and in the mid-infrared by Paladini et al. (2017) using VLTI/MIDI, showing unusually large spectral shape variations and giving evidence for brightness inhomogeneity appearing already in the dust-forming region.

The case of L$_2$ Pup is outstanding. Its light curve (Bedding et al. 2002 & 2005, McIntosh & Indermühle 2013) has been undergoing a major dimming episode in ~1994. NACO observations (Kervella et al. 2014 together with VINCI observations, Lykou et al. 2015 and Ohnaka et al. 2015 together with AMBER/VLTI observations), have revealed a dust disc, oriented east-west and inclined by ~82º with respect to the plane of the sky, with approximate projected size of 180×50 mas$^2$. Ohnaka et al. (2015) and Bedding et al. (2005) assume that dust clouds were ejected at the end of the past century, causing the formation of the dust disc. Others assume instead that it existed before the dimming event and, in 1994, simply started obscuring significantly the star. ALMA observations of continuum emission (Kervella et al. 2016) show a slight E-W elongation caused by hot dust in the disc. Finally, using SPHERE ZIMPOL observations, Kervella et al. (2015) have claimed the presence of a companion at ~2.1 au west of the star, but the NACO images and the VINCI interferometer observations do not see trace of it.

One generally finds evidence for the presence of large transparent grains very close to the photosphere, in agreement with the seminal paper of Norris et al. (2012) where VLT/NACO observations of W Hya, R Dor and R Leo were reported; in particular, dust grains surrounding W Hya, mostly aluminium composites (Takigawa et al. 2017), are found to have sizes ranging between 0.1 and 0.5 μm. Khouri et al. (2019) have detected emission from highly excited OH lines ($E_{up}$>4800 K) within a very few au from the centres of W Hya and R Dor, giving evidence for the presence of strong shock fronts.

## *3. Molecular line spectra: absorption over the stellar disc*

High angular resolution millimetre observations have revealed strong absorption of molecular line emission over the stellar disc. Wong et al. (2016) were first to give a detailed description in the case of *o* Cet. Such absorption is particularly important for SiO lines and is often found to extend beyond the stellar disc in the form of self-absorption. Absorption spectra observed over the stellar disc provide important information on the morpho-kinematics of the inner layer, in particular giving evidence for in-falling gas. They probe the central line of sight independently from maximal recoverable scale constraints and the explored radial range depends only on the radial distribution of the abundance of the molecule being studied and on the range of temperatures associated with the particular transition.

Figure 1 displays such spectra measured on the six stars studied in the present article. They show an absorption peak on the blue-shifted side, at terminal velocity, revealing the opacity of the CSE layer farther away from the star. In the case of L$_2$ Pup, where this layer is dominated by an edge-on rotating gravitationally bound disc, the absorption peak is centred at the origin. In the case of *o* Cet, where SiO molecules are confined within ~0.5 arcsec from the centre of the star, the terminal velocity is probed by the emission of CO lines but not by that of SiO lines.

With the exception of EP Aqr, all show absorption on the red-shifted side, accounting for a significant fraction of the continuum emission, usually interpreted as evidence for in-falling gas. The dominance of red-shifted absorption is qualitatively consistent with the remark that the spectra illustrated in Figure 1 are generally measured near the minimum of the light curve: the phases for *o* Cet, W Hya, R Leo and R Dor are 0.45, 0.3, 0.5 and 0.5, respectively. L$_2$ Pup is outstanding: the spectra were measured on November 5, 2015, approximately one month after the expected maximum light, which however did not occur (Kervella et al. 2016): the magnitude stayed near minimum, regular oscillations resuming in early 2016 with smaller amplitude than in 2014. Finally, there exists no light curve measurement of EP Aqr, unfortunately preventing a reliable interpretation of the absence of red-shifted absorption. As the beam is usually larger than the stellar disc, some emission from gas located farther away from us than the star is leaking into the spectrum. In case of dominant in-fall, it is seen as blue-shifted, in case of dominant out-flow, it is seen as red-shifted. In the present case, the ratio between the beam FWHM and the diameter of the

stellar disc is 13.5/12~1.1, 16/21~0.8, 23.5/26~0.9, 21/21~1, 19/32~0.6 and 16/8~2 for L$_2$ Pup, $o$ Cet, W Hya, R Leo, R Dor and EP Aqr, respectively. Its large value in the case of EP Aqr weakens the significance of absence of red-shifted absorption. All spectra show evidence for significant wings of large velocities that are discussed in Section 5.

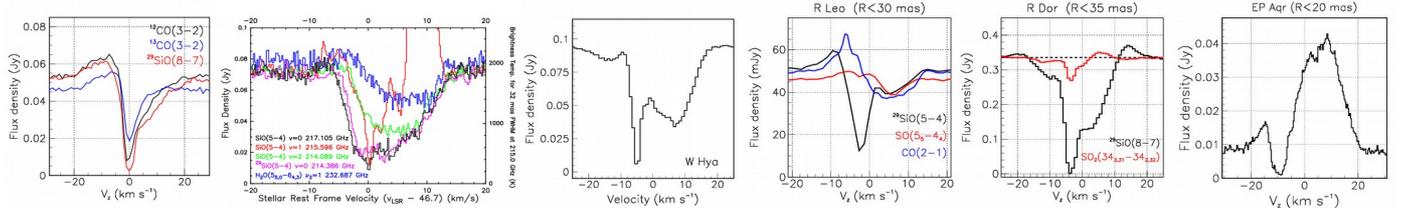

Figure 1. Doppler velocity spectra observed over the stellar disc. From left to right: L$_2$ Pup (Hoai et al. 2022a), $^{12,13}$CO(3-2) and $^{29}$SiO(8-7), 27 mas FWHM beam; $o$ Cet (Wong et al. 2016), $^{28,29}$SiO(5-4), 32 mas FWHM beam; W Hya (Hoai et al. 2022b), $^{29}$SiO(8-7), 47 mas FWHM beam; R Leo (Hoai et al. 2023), CO(2-1), $^{29}$SiO(5-4) and SO(5$_5$-4$_4$), 42 mas FWHM beam; R Dor (Nhung et al. 2021), $^{29}$SiO(8-7) and SO$_2$(34$_{3,31}$-34$_{2,32}$), 38 mas FWHM beam; EP Aqr (Nhung et al. 2023b), SiO(5-4), 32 mas FWHM beam.

## 4. Outflowing gas
### 4.1 Close to the star

Gas escaping the central gravitationally bound reservoir does so in the form of outflows covering solid angles at steradian scale and having mean radial velocities of a few km s$^{-1}$. This is a common feature observed in all six stars but taking different forms in each of them.

$o$ Cet, R Leo and R Dor display particularly complex morpho-kinematics seen as patchy emission in the close neighbourhood of the stars. On $o$ Cet (Figure 2), Nhung et al. (2022) and Hoai et al. (2020b) have used ALMA observations of SiO(5-4) and $^{12,13}$CO(3-2) line emissions to give evidence for three on-going CO outflows: one blowing south-east toward Mira B and two blowing north-east and south-west with Doppler velocities $V_z$~−3 km s$^{-1}$. Moreover, evidence is also found for recent ejections of a north-eastern arc at $V_z$~+2 km s$^{-1}$ and a south-western outflow in the plane of the sky. A soft X ray outburst was observed in 2003 (Karovska et al. 2005) using CHANDRA/ACIS-S, probably caused by a magnetic flare followed by a large mass ejection.

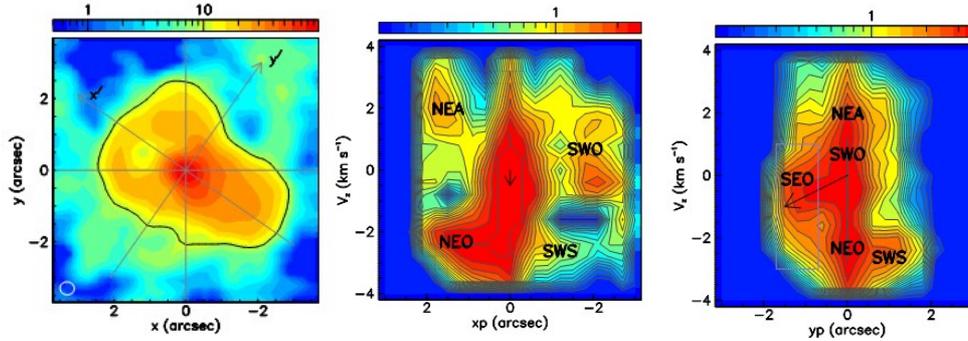

Figure 2. Outflows from $o$ Cet (Nhung et al. 2022). Left: map of the $^{12}$CO(3-2) intensity (Jy km s$^{-1}$ arcsec$^{-2}$) integrated over |$V_z$|<4 km s$^{-1}$. Middle and right: $^{13}$CO(3-2) PV maps $V_z$ vs $x'$ (middle) and vs $y'$ (right) defined in the left panel. The labels SWS, NEA, SWO and NEO stand for south-western stream, north-eastern arc, south-western outflow and north-eastern outflow, respectively; the arrow and the blue lines refer to the south-eastern outflow (SEO) dominated by the wind blowing from Mira A to Mira B. The colour scales are in units of Jy arcsec$^{-1}$.

R Leo has been studied (Hoai et al. 2023) using ALMA observations of $^{29}$SiO(5-4), CO(2-1), SO(5$_5$,4$_4$) and SO$_2$(22$_{2,20}$-22$_{1,21}$) line emissions, giving evidence for three on-going outflows covering broad solid angles in the south-eastern, south-western and north-western quadrants with low Doppler velocities at km s$^{-1}$ scale (Figure 3). The study of the relative abundance of different molecules in these outflows suggests that Local Thermal Equilibrium is violated close to the star, between ~2 and ~10 stellar radii, and is progressively restored when moving outward to ~30 stellar radii. In addition, like in $o$ Cet, detached patches of emission are observed, the one closest to the star having Doppler velocity $V_z$=2-3 km s$^{-1}$, position angle ~260° and projected distance from the star between 1 and 2 arcsec.

In the case of R Dor (Figure 4), Nhung et al. (2019) using multiline ALMA observations have shown that projected distances between ~20 and ~100 au host a radial wind with Doppler velocities reaching up to ~6 km s$^{-1}$, characterized by strong inhomogeneity, both in angle and radially, as had been noted earlier by De Beck & Olofsson (2018). Using in particular observations of the emission of the SO($6_5$-$5_4$) line, they show that the former takes the form of separate cores, of which they identify and study seven; at short distances from the star the combined effect of expansion and rotation complicates the analysis of their morpho-kinematics. Particularly outstanding is a blue-shifted stream observed at ~140° position angle to extend radially from ~10 to ~30 au (Nhung et al. 2021), which can be interpreted as a mass ejection that took place 5 to 10 years ago with a mean radial velocity of 15±5 km s$^{-1}$ but not as evidence for an evaporating planetary companion, as had been suggested by Decin et al. (2018), Homan et al. (2018a) and Vlemmings et al. (2018).

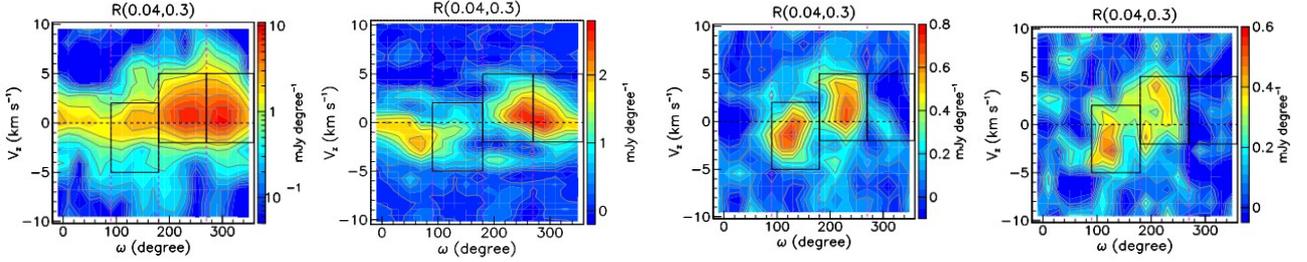

Figure 3. Outflows from R Leo (Hoai et al. 2023). PV maps, Doppler velocity $V_z$ vs position angle $\omega$, integrated over projected distances from the star 0.04<$R$<0.30 arcsec. From left to right: SiO, CO, SO and SO$_2$ emissions.

The cases of W Hya and L$_2$ Pup are different, with a single on-going outflow identified in each case (Figure 5). W Hya has been shown by Hoai et al. (2022b), using ALMA observations of $^{12}$CO(3-2) and $^{29}$SiO(8-7) emissions, to host a recent mass ejection covering a broad cone with an axis making an angle of some 30° with the line of sight, and a blob of enhanced emission located some 0.2 arcsec (20 au) above the plane of the sky. Within the cone, and within radial distances confined to less than 25 au from the centre of the star, the number density is increased by an order of magnitude, with a SiO/CO ratio larger by at least a factor 2, and the effective line width is also about twice as large. The mean radial velocity inside the cone is 6-7 km s$^{-1}$. The cone geometry matches the dust distribution observed by Ohnaka et al. (2017) using SPHERE/ZIMPOL at the VLT.

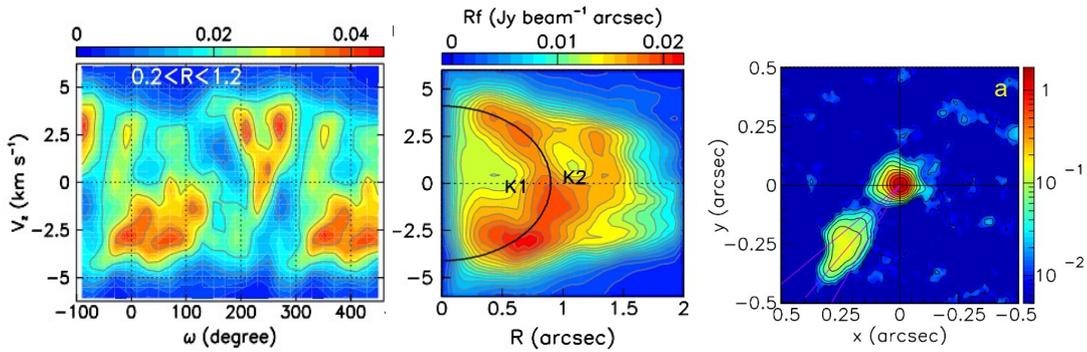

Figure 4. Outflows from R Dor, SO($6_5$-$5_4$) emission (Nhung et al. 2019). Left: P-V map (Doppler velocity $V_z$ vs position angle $\omega$) integrated over projected distances from the star 0.2<$R$<1.2 arcsec. The colour scale is in units of Jy beam$^{-1}$. Middle: P-V map of the product brightness×angular distance in the $V_z$ vs $R$ plane, averaged over $\omega$, showing toroidal cavity K$_1$ (and possibly K$_2$), giving evidence for a recent episode of enhanced mass-loss. Right: the blue-shifted stream seen in $^{29}$SiO(8-7) emission (Nhung et al. 2021). The intensity map is integrated over −18.7<$V_z$<−6.3 km s$^{-1}$.

Close to the star, the CSE of L$_2$ Pup is dominated by a gas-and-dust disc in rotation, nearly edge-on. It may be slightly expanding at a rate of 0.8±0.5 km s$^{-1}$. It has been studied using ALMA observations of $^{29}$SiO(8-7) emission (Kervella et al. 2016), of $^{12,13}$CO(3-2) emission (Homan et al. 2017) and $^{29}$SiO(8-7), $^{12,13}$CO(3-2) and $^{28}$SiO(5-4) emissions (Hoai et al. 2022a). An on-going northern outflow with $V_z$ between −5 and −3 km s$^{-1}$ is visible in $^{12}$CO and $^{28}$SiO emissions and probably associated with a feature seen in NACO infrared images (Kervella et al. 2014); it is the only candidate mass ejection that can be clearly identified (Hoai et al. 2022a).

The case of EP Aqr is outstanding (Figure 6). In addition to two broad outflows observed in SiO(5-4) line emission, of a similar nature as those observed in the other stars of the sample, a disc in rotation, displaying emission dominated by CO lines, is observed to be formed very close to the photosphere. While Homan et al. (2018b, 2020a) have proposed an interpretation in terms of a white dwarf companion orbiting the star at an angular distance of ~400 mas, the currently favoured interpretation (Nhung et al. 2023b) relies on a different mechanism for the formation of the disc and interprets the observed morpho-kinematics of the CSE and the apparent formation of polar outflows as the result of the interaction between the disc and a standard stellar wind.

In all cases, the radial extension of the outflows and the distance between separated patches are at the scale of some 100 au, a distance covered in 50 years at a velocity of 10 km s$^{-1}$. Such a time scale is more than two orders of magnitude larger than the typical variability observed on the stellar disc. Whatever is causing these outflows to blow in a same direction over several years must have a lifetime at the same scale.

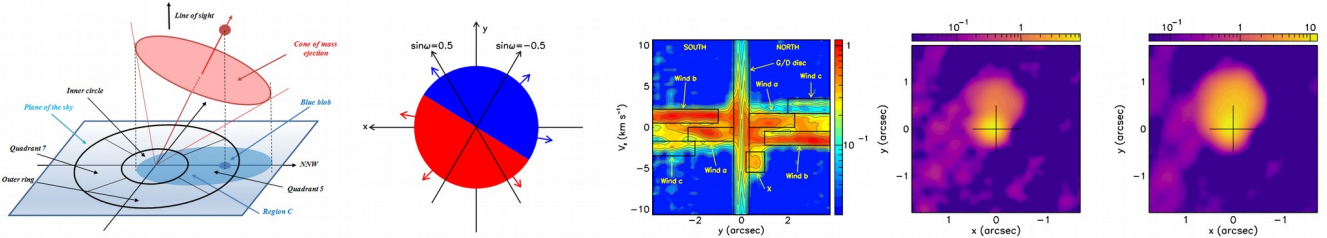

Figure 5. Outflows from W Hya (left, Hoai et al. 2022b) and L$_2$ Pup (right panels, Hoai et al. 2022a). Left: schematic geometry of the mass ejection from W Hya. Middle pair of panels: $^{12}$CO(2-1) emission of L$_2$ Pup. The left panel shows a schematic view ($x$ pointing east and $y$ pointing north) of the expanding dominant $b$ wind, inclined in the NW/SE direction with respect to the plane of the sky, and of the wedge, defined as $|\sin\omega|<0.5$, within which the PV map shown in the right panel is integrated. The PV map is in the $V_z$ vs $y$ plane and indicates the various components that contribute to the emission. The central gas-and-dust disc is indicated as G/D disc and the on-going outflow mentioned in the text is indicated as X. The colour scale is in units of Jy arcsec$^{-1}$. Right pair of panels: intensity maps of $^{12}$CO(2-1) (left) and $^{28}$SiO(5-4) (right) emissions integrated over $-5<V_z<-3$ km s$^{-1}$, showing the emission of blob X defined in the central pair of panels.

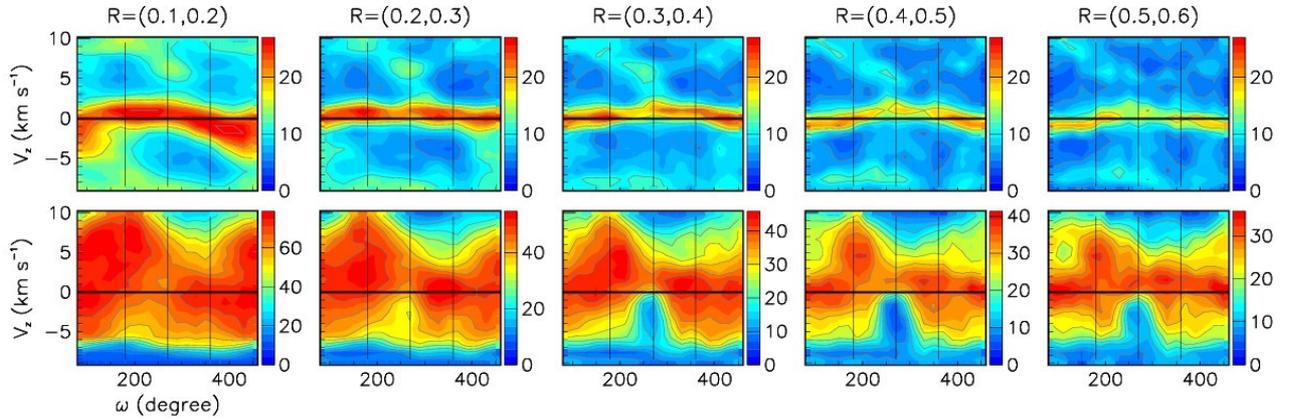

Figure 6. EP Aqr (Nhung et al. 2023b). PV maps (Doppler velocity vs position angle) integrated over five successive intervals of angular distance as indicated on top of the panels in arcsec. In the upper row CO(2-1) emission is dominated by a disc, in the lower row SiO(5-4) emission is dominated by a pair of outflows.

*4.2 Farther away*

A similar comment applies to the CSE's at larger distances, where the history of the mass-loss mechanism is explored at century scale. They are mostly studied using ALMA observations of the emission of CO lines, most SiO molecules having either been dissociated by UV radiation or adsorbed on dust grains. While covering the whole 4π solid angle, they display strong inhomogeneity and deviate significantly from spherical envelopes, taking rather the form of radially expanding oblate volumes more or less inclined with respect to the plane of the sky and hosting significant localized emission enhancements. Such are the cases of R Dor (Homan et al. 2018a, Hoai et al. 2020a, Nhung et al. 2019 & 2021), L$_2$ Pup (Hoai et al. 2022a) and possibly $o$ Cet (Hoai et al. 2020b, Nhung et al. 2022). The oblate CSE of R Dor (Figure 7) displays enhanced emissions in two back-to-back directions, each dominated by a pair of cores. The CSE of $L_2$ Pup probes the mass-loss history preceding the 1994 dimming event and hosts three directions of enhanced emission. The CSE of EP Aqr (Homan et al. 2018b, 2020a, Hoai et al. 2019) is characterised by broad bipolar outflows along the disc axis, the two-component morpho-kinematics being clearly revealed by the Doppler velocity spectrum (Figure 8); the currently favoured interpretation assumes that it is the interaction of a standard wind with the equatorial disc that causes the wind expansion velocity to decrease from ~10 km s$^{-1}$ on the disc axis down to ~2 km s$^{-1}$ in the disc plane, producing the appearance of two polar outflows (Nhung et al. 2023b). Very patchy emission is observed in the CSEs of R Leo and $o$ Cet (Figure 9). In the cases of R Leo (Hoai et al. 2023) and R Dor (Nhung et al. 2019), episodes of enhanced mass-loss are revealed by toroidal depressions of the observed emission. The observed terminal velocities (Table A1) never exceed ~12 km s$^{-1}$, reaching such a value in the only case of the bipolar outflows of EP Aqr, for which the terminal velocity of the equatorial density enhancement is only 2 km s$^{-1}$; the mean terminal velocity of the five other stars is 5.4 km s$^{-1}$ with an rms deviation of only 0.8 km s$^{-1}$ with respect to the mean.

It is tempting to infer from these observations that whatever is causing ejection of gas from the extended atmosphere of the star, as do probably shocks induced by pulsations and convective cells, while displaying rapid variations at the scale of months, retain a same broad deviation from sphericity at century scale. However, such conclusion is weakened by the remark that other mechanisms contribute to shaping the CSE away from the star. Such is particularly the case of the presence of stellar or planetary companions. In the present sample, the case of EP Aqr is the most compelling (Nhung et al. 2023b), the presence of polar outflows being understood as resulting from effective collimation by the continuous formation of a slowly expanding disc.

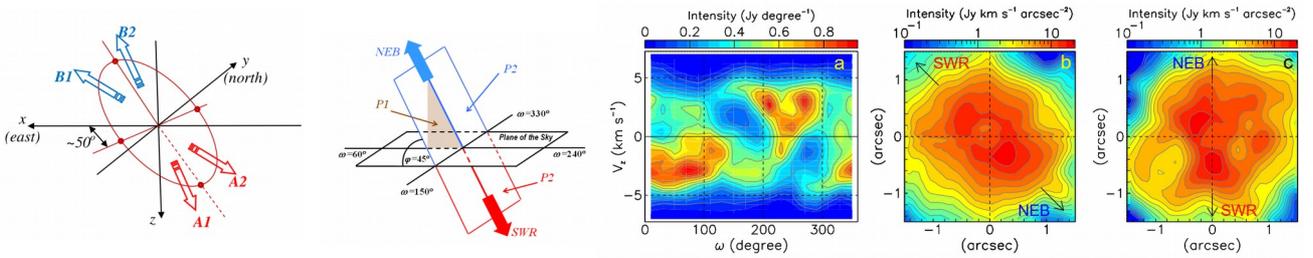

Figure 7. R Dor CSE. Left panel: schematic of the morpho-kinematics (Nhung et al. 2019). Right panels, SO($6_5$-$5_4$) line emission (Nhung et al. 2021). From left to right: definition of the geometry and of planes P$_1$ and P$_2$, (a): intensity in the Doppler velocity $V_z$ vs position angle $\omega$ plane integrated over 12 au<$R$<60 au, (b): intensity projected on P1; (c): intensity projected on P2. In both panels (b) and (c) NEB stands for North-east-blue-shifted and SWR for South-west-red-shifted and a cut $R$>0.3 arcsec (18 au) has been applied.

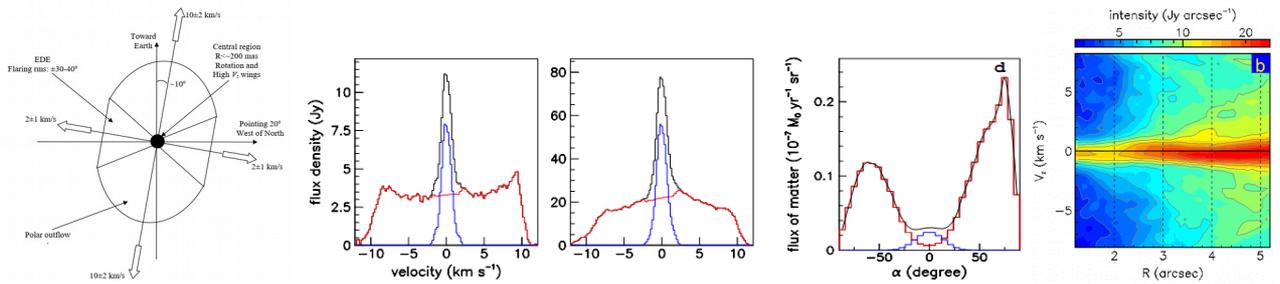

Figure 8. EP Aqr. Left: schematics of the morpho-kinematics of the CSE. Central triplet of panels (Hoai et al. 2019): two-component structure of the Doppler velocity spectra of $^{13}$CO(3-2) (left) and $^{12}$CO(3-2) (middle) emissions, and dependence of the flux of matter on stellar latitude (right). The narrow component (equatorial density enhancement) is shown in blue and the broad component (bipolar outflow) in red. Right: PV map in the $V_z$ vs $R$ plane of the projection of the data cube defined as 1.2<$R$<5.2 arcsec and |$V_z$|<8 km s$^{-1}$.

## 5. High Doppler velocity wings and possible rotation in the inner CSE layer
### 5.1 General comments

All six stars give evidence for significant effective line broadening, in particular of SiO lines, at distances from the star below ~15 au, the less convincing case being of W Hya, for which clear line broadening is only seen over the stellar disc. It is observed in the form of high Doppler velocity wings above continuum level and confined to the vicinity of the line of sight crossing the star at its centre; they extend to some ±10-20 km s$^{-1}$ in velocity and to some 15 au in radius. They receive negligible contributions from thermal broadening and are understood as being in fact the result of the simultaneous observation of high-velocity blue-shifted and red-shifted gas flows on an angular scale smaller than the beam size. We recall that the maximal distance reached by a volume of gas launched with velocity $V_0$ at distance $r_0$ is $r_{max}=r_0V_{esc}^2/(V_{esc}^2–V_0^2)$ where the escape velocity $V_{esc}=(2GM/r_0)^{½}$, $G$ being Newton's gravity constant and $M$ the mass of the star. Namely for a radial boost of velocity at the star photosphere equal to a fraction $f$ of the escape velocity, it will reach $1/(1–f^2)$ stellar radii: it remains below ~5 stellar radii for $f<0.9$ and then blows up rapidly when $f$ approaches unity. For a solar mass star of radius 1 au, the escape velocity is 42 km s$^{-1}$ and a 20 km s$^{-1}$ boost will therefore produce a maximal distance of ~1.3 stellar radii.

The presence of high velocity wings had been noticed early in single dish observations (Winters et al. 2002), suggesting that the emission occurs close to the star, where SiO grains have not yet fully formed, and is somehow related to star pulsations. Their ALMA observation was first mentioned by Decin et al. (2018) and early attempts at interpreting them as bipolar gas streams flowing along the line of sight in the case of EP Aqr (Tuan-Anh et al. 2019) or as the effect of rotation in the case of L$_2$ Pup (Kervella et al. 2016 and Homan et al. 2017) have been published. However, it has been later understood, following the study of other stars, that they are in fact confined to the close neighbourhood of the star and are essentially radial, although occasionally combining with rotation (Nhung et al. 2023a & 2023b). As first remarked by Decin et al. (2018), it is the high sensitivity of ALMA observations, rather than their high angular resolution, that made it possible to reveal their presence. Indeed, to the extent that their source is confined to the close neighbourhood of the star and that no other source of such high Doppler velocities exists, low angular resolution observations are able to detect them as long as their sensitivity is sufficient: their emission is a minor fraction of the total emission detected within the broad beam, but they have no competition in the extreme Doppler velocity region that they populate. That does not mean that high angular resolution observations are unnecessary. They are indeed essential to measure the radial extension of the source and to explore the details of its morphology projected on the plane of the sky.

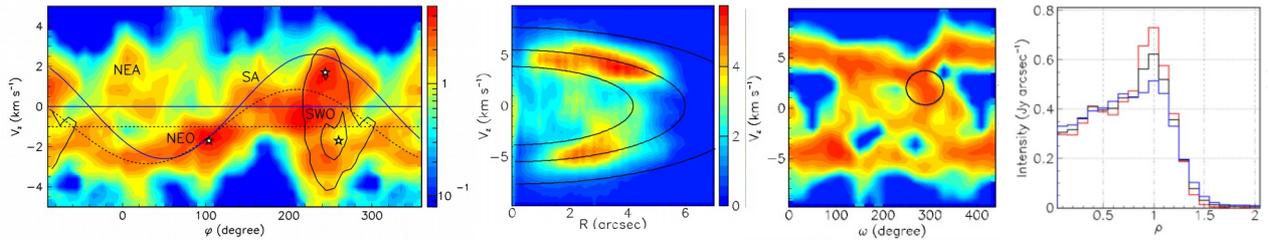

Figure 9. Left panel: the *o* Cet emissions of the CO(3-2) (colour) and SiO(5-4) (contours) lines integrated over 1.0<*R*<3.7 arcsec are shown in the Doppler velocity $V_z$ vs position angle $\varphi$ plane (Hoai et al. 2020b). The colour scale is in Jy. NEA, SWO, SA and NEO stand for North-eastern arc, South-western outflow, Southern arc and North-eastern outflow, respectively. The full (dashed) sinusoidal lines are the traces of an isotropic radial wind of 2.8 (2.0) km s$^{-1}$ velocity blowing in the plane of Mira B orbit using systemic velocities of 47.7 (46.7) km s$^{-1}$. Right triplet of panels: R Leo emission of the CO(2-1) line (Hoai et al. 2023) integrated over solid angle and displayed in the Doppler velocity vs *R* plane (left) or integrated over 2<*R*<3 arcsec and displayed in the Doppler velocity vs position angle plane (middle). The lines on the left panel show ellipses $\rho=1/\sqrt{2}$, 1 and $\sqrt{2}$ with $\rho=([V_z/5.5 \text{ km/s}]^2+[R/6"]^2)^{½}$. The right panel displays the dependence on $\rho$ of the intensity integrated over the red-shifted (red) and blue-shifted (blue) hemispheres and averaged over velocity channels.

Even with an excellent angular resolution, the observed line profile receives contributions from all sources located on the line of sight; the opacity of the layers crossed at the line frequency suppresses part of their emission but, at extreme Doppler velocities, the density can be expected to be low enough for absorption to be small. Several effects contribute to the dependence of the line emission on radial distance: depletion of the gas phase by adsorption on dust grains, as is the case for SiO and AlO; mode of excitation, by collision with other molecules, mostly hydrogen, or absorption of stellar UV radiation; photo-dissociation by interstellar radiation or by radiation emitted by the star itself or by a nearby white dwarf companion; dissociations produced by possible shocks; dependence on temperature, the relevant parameters being the emission parameter $\varepsilon$, which measures the emissivity in the absence of absorption, and the optical depth $\tau$.

Absorption, particularly strong over the stellar disc but also extending well beyond it in some cases (see Section 3), complicates the study of line broadening, the line profile being interrupted over a significant interval of Doppler velocity. Its contribution needs to be taken in proper account. The complexity of the physics mechanisms at stake is particularly well illustrated by the analyses of Wong et al. (2016) for $o$ Cet and Vlemmings et al. (2017) for W Hya.

Evidence for rotation in the inner region of the CSE, at distances from the star comparable with those covered by significant effective line broadening, has been found in four stars of the present sample, particularly strong in $L_2$ Pup and R Dor, but absent in W Hya and R Leo. The presence of rotation in the same radial range as line broadening requires disentangling the two effects, which can be done by considering separately four quadrants of position angle: near the projection of the rotation axis on the plane of the sky, the effect of rotation cancels, while it is maximal in the perpendicular direction. Similarly, the possible presence of expansion in this radial range complicates the analysis of the line broadening and needs to be properly taken in account. A detailed discussion of disentangling the effects of line broadening, rotation and expansion has been given by Nhung et al. (2023a) using the case of R Hya as an illustration.

## 5.2 Selected sample
### 5.2.1 $L_2$ Pup

Hoai et al. (2022a) and Homan et al. (2017) have given evidence for high $|V_z|$ wings in the $^{29}$SiO(8-7) and $^{12,13}$CO(3-2) emissions of the inner CSE of $L_2$ Pup: over ~60 mas from the centre of the star, the effective line width increases (FWHM) from ~5 km s$^{-1}$ to ~20 km s$^{-1}$ (Figure 10, upper row). The rotation of the gas and dust disc is seen on all observed lines (Figure 10, lower row) with a projected rotation velocity increasing from ~4 km s$^{-1}$ to ~8 km s$^{-1}$ (Figure 10, upper row) when $R$ decreases from [120-150] mas to [30-60] mas (Hoai et al. 2022a). However, ignoring line broadening, Kervella et al. (2016) claim instead a projected rotation velocity of ~15 km s$^{-1}$ in the [30-60] mas interval.

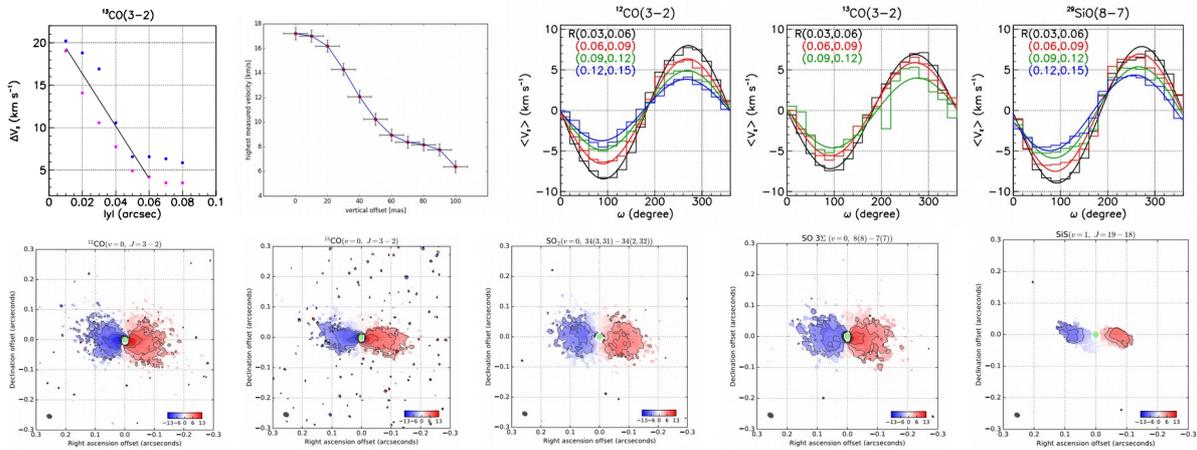

Figure 10. Upper row, left pair of panels: line broadening in $L_2$ Pup measured on the $y$ axis where contribution from rotation cancels as seen from $^{13}$CO(3-2) line emission (left, Hoai et al. 2022a) and $^{12}$CO(3-2) line emission (right, Homan et al. 2017); the left panel shows the evolution of the FWHM line width (blue and red correspond to south and north, respectively) and the right panel the maximal velocity. Upper row, right triplet of panels: $L_2$ Pup (Hoai et al. 2022a), dependence on position angle of the Doppler velocity averaged over $|V_z|$<20 km s$^{-1}$ in successive rings as indicated in the inserts (in arcsec); the lines show sine wave fits. Lower row: maps of the mean Doppler velocity measured for several $L_2$ Pup lines observed with high angular resolution by Kervella et al. (2016).

### 5.2.2 $o$ Cet
Hoai et al. (2020b) and Nhung et al. (2022) give strong evidence for the line width of the SiO emission to increase from $|V_z|$~5 km s$^{-1}$ to ~20 km s$^{-1}$ when $R$ decreases from ~0.17 to ~0.05 arcsec and for rotation about an axis projecting 40°±15° east of north with a projected velocity of 0.7±0.2 km s$^{-1}$, in agreement with results obtained from the observation of SiO masers (Cotton et al. 2004, 2006). The low rotation velocity and the absence of absorption outside the stellar disc make this case relatively easy to analyse (Figure 11).

5.2.3 W Hya

The study of line broadening in W Hya (Vlemmings et al. 2017, Hoai et al. 2022b) is made difficult by the presence of a recent mass ejection in the blue-shifted northern octant; outside the region of the sky plane covered by this mass ejection, observations of the $^{29}$SiO(8-7) line emission using a 45 mas FWHM beam show that the largest Doppler velocities are observed within ~75 mas projected distance from the centre of the star and reach only ~10 km s$^{-1}$, less than twice the terminal velocity. The clearer evidence for larger Doppler velocities is from the absorption spectra measured over the stellar disc, reaching ~± 20 km s$^{-1}$ and showing evidence for in-falling gas reaching high Doppler velocity. Vlemmings et al. (2017), using a beam of only ~16 mas FWHM, have observed the CO($v$=1,3-2) line emission of W Hya, which, with an upper level at 3120 K, probes efficiently the warm extended atmosphere. Their analysis reveals the presence of a warm molecular layer close to the stellar surface, with temperatures nearing 3000 K, surrounded by a cooler layer at ~900 K, extending to radial distances of ~50 mas and hosting both in-fall and outflow components (Figure 12). These reach larger velocities in the eastern hemisphere than in the western hemisphere where a hot spot of continuum emission is observed. No evidence for rotation has been reported.

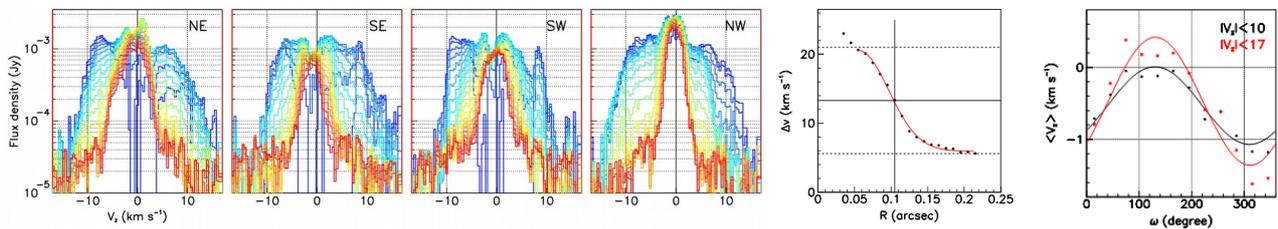

Figure 11. Line broadening and rotation in the SiO(5-4) line emission of *o* Cet (Hoai et al. 2020b and Nhung et al. 2022). Left quartet of panels: Doppler velocity spectra averaged in quadrants of 3 au broad annular rings centred on Mira A with mean radii increasing from 2.5 au (blue) to 21.5 au (red) in steps of 1 au. The dependence on *R* of the full-width at 1/5 maximum of the spectra is shown in the next panel. Right: dependence of the mean Doppler velocity on position angle *ω* in the ring 3 au<*R*<10 au. The lines are sine wave best fits.

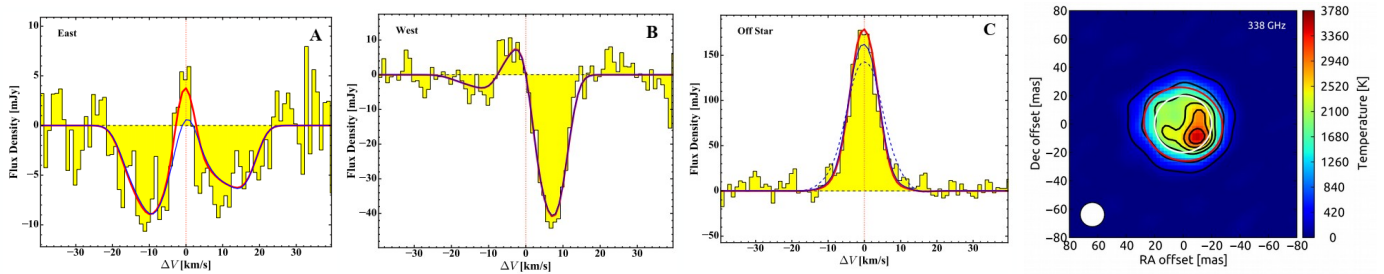

Figure 12. Doppler velocity spectra of the CO($v$=1,3-2) line emission of W Hya (Vlemmings et al. 2017) extracted towards the eastern (A) and western (B) stellar hemisphere as well as in an annulus between 2.9 and 5 au off the star (C). The beam size is ~16 mas FWHM and the spectral resolution ~1 km s$^{-1}$. The red lines show model results. The right panel maps the brightness temperature of the stellar surface as observed with ALMA continuum emission at 338 GHz. The rms noise in the image is 18 K and the peak is 3560 K. The red ellipse indicates the size of the fitted, uniform, stellar disc with a brightness temperature of 2495±125 K. The effective temperature of the photosphere is ~2500 K.

5.2.4 R Leo

In the case of R Leo, Hoai et al. (2023) observe Doppler velocities reaching beyond 10 km s$^{-1}$ over the stellar disc for the $^{29}$SiO(5-4), CO(2-1) and SO($5_5$-$4_4$) lines. They decrease to ~10 km s$^{-1}$ for *R* between 30 and 50 mas and reach terminal values of ~5 km s$^{-1}$ for *R* beyond ~100 mas. They show important anisotropy, with stronger emission in the southern than northern hemisphere. Fonfría et al. (2019a) have observed the emission of the CO(2-1) and $^{29}$SiO(5-4) lines over the stellar disc (Figure 13) giving evidence for in-falling gas in front of the star and for most of the emission coming from directions that do not enclose the star; they comment on the possibility for the $^{29}$SiO($v$=1,6-5) line to show maser emission. They argue that the red-shifted absorption may be produced by radial movements of photospheric layers (or convective cells) due to the stellar pulsation and that the spectra suggest the presence of random gas movements in the stellar photosphere and/or of periodic shocks. They also suggest the possible presence of rotation, which however cannot be confirmed when the outflows described by Hoai et a. (2023) are taken in due account. Higher angular resolution observations would be necessary to better explore the close environment of R Leo.

5.2.5 R Dor
In the case of R Dor, the presence of high Doppler velocity wings projecting on nearly central lines of sight and reaching 15-20 km s$^{-1}$ had been noted early by Decin et al. (2018) and Hoai et al. (2020a) (Figure 14); Nhung et al. (2021), using observations of the line emissions of $^{29}$SiO (8-7) and SO$_2$(34$_{3,31}$-34$_{2,32}$) and accounting for the contribution of rotation, have shown that such significant line broadening occurs within ~12 au from the centre of the star. In this region evidence for rotation was first obtained by Vlemmings et al. (2018) using high angular resolution observations of SiO($v$=3,5-4), SO$_2$(16$_{3,13}$-16$_{2,14}$) and $^{29}$SiO($v$=1,5-4) emissions and, farther away from the star, by Homan et al. (2018a) using observations of the SiO($v$=1,8-7) emission. The former authors gave evidence for solid-body rotation with velocity increasing from ~1 km s$^{-1}$ near the stellar surface and the latter authors for sub-Keplerian rotation with velocity cancelling beyond 15 au from the centre of the star. Nhung et al. (2019, 2021) have later shown that rotation reaches a maximal velocity of ~5 to 7 km s$^{-1}$ at some 8±2 au from the centre of the star, where the Keplerian orbital velocity around a solar mass star is ~11 km s$^{-1}$. Its axis projects on the plane of the sky at a position angle of 10±5º and makes an unknown angle with the plane of the sky; yet, qualitative considerations favour a low value, typically 20º±20º as suggested by the analysis of Homan et al. (2018a). The morphology of the rotating volume does not show strong anisotropy and, in particular, fails to provide evidence for a disc-like flattening about the equator as observed in the case of L$_2$ Pup.

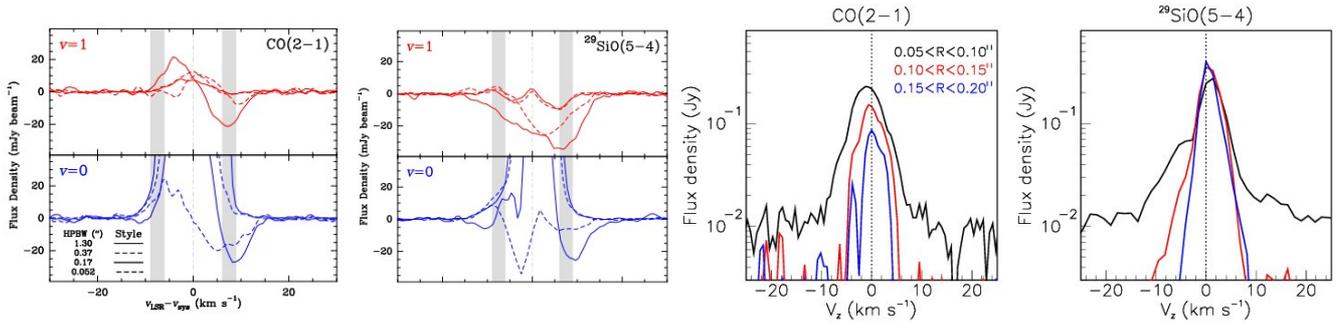

Figure 13. High velocity wings in R Leo. Left pair of panels: Fonfría et al. 2019a, Doppler velocity spectra in the central pixel of the $^{12}$CO(2-1) and $^{29}$SiO(5-4) lines ($v$=0 in blue and $v$=1 in red). The different line styles correspond to different beam sizes as indicated in the insert. The grey vertical bands indicate the gas expansion velocity derived from single-dish observations. Right pair of panels: Hoai et al. 2023, Doppler velocity spectra of the CO(2-1) and $^{29}$SiO(5-4) line emissions in 50 mas wide rings as indicated in the insert. Note that the continuum has not been subtracted.

5.2.6 EP Aqr
High Doppler velocity wings, reaching ~±20 km s$^{-1}$, have been observed on both SiO(5-4) and CO(2-1) line emissions and shown to be confined within ~150 mas from the centre of the star (Nhung et al. 2023b) (Figure 15). The simultaneous presence of large radial wind velocities, reaching above 12 km s$^{-1}$ complicates their study. Evidence for rotation within some 100-200 mas from the centre of the star, with a rotation velocity at km s$^{-1}$ scale, has been obtained from the observation of the emissions of the CO(2-1) and SO$_2$(16$_{6,10}$-17$_{5,13}$) lines, consistent with observations of the dust induced polarization of the 550-750 nm emission observed at the VLT using SPHERE-ZIMPOL (Homan et al. 2020a). This suggests that such rotation may be related to the precocious formation of the equatorial disc. However, the orientation of the disc, inclined by only ~10º with respect to the plane of the sky, makes its study particularly difficult and additional high sensitivity and angular resolution observations are necessary to clarify this point.

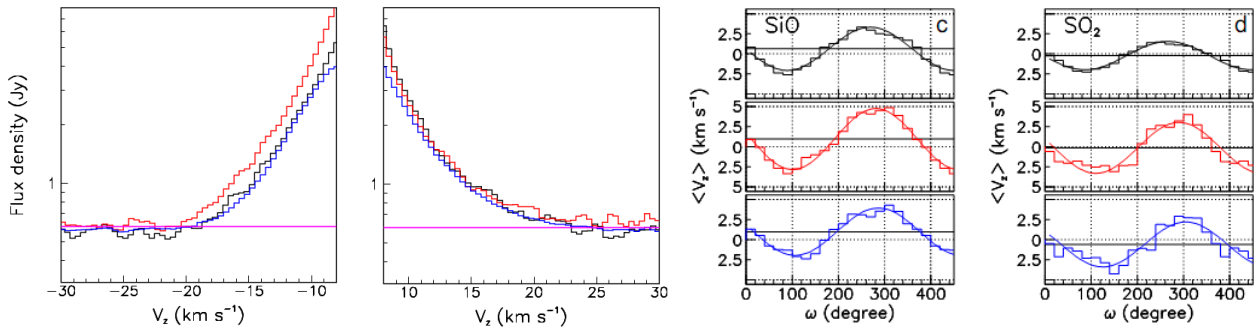

Figure 14. R Dor inner CSE layer (Nhung et al. 2021). Left pair of panels: Doppler velocity distributions of the $^{29}$SiO(8-7) emission observed for $|V_z|$ >8 km s$^{-1}$ using different datasets. Horizontal lines show the continuum level. Right pair of panels: dependence on position angle of the mean Doppler velocity averaged over 50<$R$<100 mas (black), 100<$R$<150 mas (red) and 150<$R$<200 mas (blue) rings for $^{29}$SiO(8-7) and SO$_2$(34$_{3,31}$−34$_{2,32}$) line emissions.

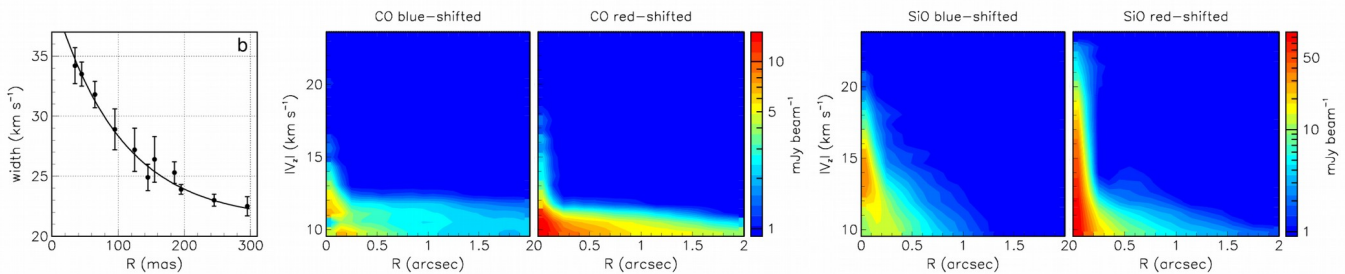

Figure 15. EP Aqr. Left: radial dependence of the full width at 10% of maximum of the Doppler velocity spectrum measured in a ring of radius $R$ centred on the star. Four rightmost panels: PV maps of $|V_z|$ vs $R$ for both hemispheres and both CO(2-1) and SiO(5-4) lines separately as indicated on top of each panel.

## 6. Discussion and conclusion
### *6.1 Rotation and the role of planetary or stellar companions*
Evidence for rotation in the close neighbourhood of the star is generally ascribed to the presence of planetary or stellar companions. Indeed, while it is recognized that the initial angular momentum of the star must produce negligible velocities on the surface of the AGB star, which has expanded by over two orders of magnitude, planetary companions, whether orbiting the star above the photosphere or having been recently engulfed (Privitera et al. 2016), can be expected to transfer a significant part of their angular momentum to the inner CSE. In contrast to protostars, the complex dynamics at stake near the photosphere is likely to prevent the radial distribution of the velocity of such rotating gas from being Keplerian. While some carbon-rich evolved stars, such as LL Pegasi (Morris et al. 2006), display very clear spiral features revealing the presence of a companion, oxygen-rich stars do not. Attempts at identifying such structures have always proven difficult. A well-documented recent example was given by Randall et al. (2020) in the case of GX Monocerotis. In the particular sample of nearby stars considered in the present article, claims for the presence of spiral structures have been made by Ramstedt et al. (2014) in the case of *o* Cet and by Homan et al. (2018b, 2020a) in the case of EP Aqr. However, the former has been shown by Hoai et al. (2020b) to wind in the wrong direction and serious doubts have been shed by Nhung et al. (2023b) on the reality of the latter. While the validity of the observation of possible spiral structures must be critically assessed, the failure to obtain reliable evidence should not be understood as a denial of the presence of planetary companions, which are obviously expected to be there. The question, which has currently received no satisfactory answer, is of their precise role in the generation of the wind.

In the present sample, the only star, which has a well-established companion, in this case probably a white dwarf (Mira B), is *o* Cet. It plays a clear, but modest, role in focusing the wind of the main star (Hoai et al. 2020b, Nhung et al. 2022).

The case of EP Aqr is outstanding. The evidence for the precocious formation of a nearly face-on equatorial disc is strong, the disc interacting with the nascent wind, giving the CSE its two-component appearance, with a pair of polar outflows. In such a case, the question to be answered is not the nature of the generation of the wind, assumed to obey the current state-of-the art ideas developed in the Standard Model, assumed to obey current state-of-the art ideas (Höfner & Freytag 2019), but the mechanism that governs the formation of the disc, which is unknown but might very well imply the presence of a planetary companion near the photosphere.

In the present sample, the two stars, which display clearer evidence for significant rotation, are R Dor and $L_2$ Pup. In the former case (Nhung et al. 2021, Vlemmings et al. 2018, Homan et al. 2018b), rotation reaches a maximal velocity of ∼6 km s$^{-1}$ at some 4 stellar radii from the centre of the star, and cancels beyond ~8 stellar radii, suggesting the presence of a planetary companion close to the photosphere or having been recently engulfed. The lack of significant oblateness disproves the presence of a narrow disc. There is no apparent relation between the geometry of the rotation (an axis projecting ~10° east of north on the plane of the sky) and that of the median plane of the global CSE (with a normal projecting ~60° east of north), probably implying that the observed rotation does not play an important role in the formation of the nascent wind.

The case of $L_2$ Pup (Kervella et al. 2016, Homan et al. 2017, Hoai et al. 2022a) is different: the rotating volume is a nearly edge-on disc of gas and dust covering radial distances between 30 and 400 mas from the centre of the star and rotation velocities reaching ~8 km s$^{-1}$ near the inner rim (the latter evaluation taking proper account of the effective line broadening in the vicinity of the star). As a result, at strong variance with the disc of EP Aqr, the velocities are accurately measured but the radial distribution is not. This makes it difficult to compare the two discs. The $L_2$ Pup disc seems, however, to be of a different nature: its outer diameter is much smaller, it hosts line emissions from both CO and SiO molecules while the EP Aqr disc is formed near the photosphere, where it is revealed exclusively by the emission of CO molecules and extends to large distances; in contrast with the EP Aqr disc, the $L_2$ Pup disc does not seem to play a role in shaping the CSE: farther away from the star, the wind oblateness is unrelated to the orientation of the disc and no evidence has been found for outflows collimated by the disc. This implies that for $L_2$ Pup, as for R Dor, the observed rotation does not play an important role in the formation of the nascent wind. But understanding the mechanism that governed the formation of the rotating disc is complicated by the presence of the strong dimming episode that the light curve underwent at the end of the past century, inviting different speculations. Yet, it is generally assumed that an orbiting planetary companion is responsible for the observed rotation and Kervella et al. (2016) claim to have obtained evidence for it, although their analysis suffers from ignoring effective line broadening.

In summary, while rotation is present within a few stellar radii of the centre of several oxygen-rich AGB stars, there is no evidence for it to play an important role in the generation of the nascent wind. There is no reason to doubt the presence of planetary companions orbiting AGB stars at short distances from their centre or having been recently engulfed, but their role seems limited to their impact on shaping the gravity field and to the production of rotation by dragging gas, with no significant incidence on the generation proper of the nascent wind. When based on the observation of an apparent spiral wake, claiming evidence for the presence of a companion requires a critical analysis, sufficiently rigorous to be credible. Cases where the presence of a companion is well established, such as *o* Cet in the present sample or R Aqr (Schmid et al. 2017, Bujarrabal et al. 2018 & 2021) and $\pi 1$ Gru (Homan et al. 2020b) may display spectacular symbiotic phenomena, but the role of the companion is limited to its impact on shaping a pre-existing wind, with no incidence of its preliminary formation. Yet, more detailed studies of the mechanism governing the formation of rotation in the inner layers of the CSE, complicated by the presence of effective line broadening, are required in order to better specify the role played by possible companions.

***6.2 Effective line broadening and the role played by stellar pulsations and convective cell granulation***
The commonly accepted ideas of the Standard Model about the mechanisms governing the formation of the nascent wind of oxygen-rich AGB stars, outlined in the Introduction, are proving very useful in guiding current research and helping with the interpretation of the results. They have not suffered any significant contradiction. Yet, their validity is far from being confirmed and their conformity with observations is mostly qualitative.

Indeed, the present ALMA observations identify the region where the wind gets its initial boost with the radial range over which effective line broadening is observed. The complexity of the morpho-kinematics at stake in this region is described by essentially just two numbers: its mean distance from the centre of the star, $r_{boost}$, and the extreme velocity reached by the high $V_z$ wings, $V_{boost}$. Table 1 lists their values together with the terminal velocity $V_{term}$, the stellar radius $r_{star}$ and the mass-loss rate, $\dot{M}$. The values and uncertainties retained for W Hya, $V_{boost}$=15±5 km s$^{-1}$ and $r_{boost}$=40±30 mas reflect the difficulty to summarize in only two numbers the complex morpho-kinematics revealed by Vlemmings et al. (2017) and Hoai et al. (2022b). As illustrated in Figure 16, the values of $V_{boost}$ are similar, with a mean value of 15.3 km s$^{-1}$ and an rms deviation of 2.6 km s$^{-1}$ with respect to the mean. The case of EP Aqr is outstanding, with a value of $V_{term}$ (12±2 km s$^{-1}$) more than twice that of the other stars of the sample (5.4±0.8 km s$^{-1}$) and a ratio $r_{boost}/r_{star}$ ~12, four times as large, resulting from a very small stellar radius, ~7 mas, and a relatively broad line broadening region, reaching over 150 mas from the centre of the star. For the other stars, $r_{boost}$ increases in near proportion to $r_{star}$, with a ratio $r_{boost}/r_{star}$~3; it is higher for $o$ Cet, probably because of the larger mass of the star causing the mass-loss rate to be small in spite of the large value of $r_{boost}$. L$_2$ Pup, with the smallest values of both $r_{boost}$ and $\dot{M}$, stands out as being particularly quiet. The crudeness of these considerations prevents us from making deeper comments and illustrates well the oversimplification implied by such description.

Table 1. Parameters of relevance to the region of effective line broadening.

|  | *L$_2$ Pup* | *o Cet* | *W Hya* | *R Leo* | *R Dor* | *EP Aqr* |
|---|---|---|---|---|---|---|
| $V_{boost}$ *(km s$^{-1}$)* | *17±3* | *12±2* | *15±5* | *13±3* | *15±3* | *20±2* |
| $V_{term}$ *(km s$^{-1}$)* | *4±2* | *5±2* | *6±2* | *6±2* | *6±1* | *12±2* |
| $r_{boost}$*(mas)* | *35±10* | *105±15* | *40±30* | *50±20* | *100±20* | *85±15* |
| $r_{star}$ *(mas)* | *11±2* | *18±3* | *23±3* | *17±4* | *30±2* | *7±2* |
| *d (pc)* | *~64* | *~100* | *~104* | *~114* | *~59* | *~119* |
| $\dot{M}$ *(10$^{-7}$M$_{Sun}$/yr)* | *0.12±0.07* | *0.7±0.2* | *1.3±0.2* | *1.6±0.2* | *2±0.2* | *1.6±0.4* |

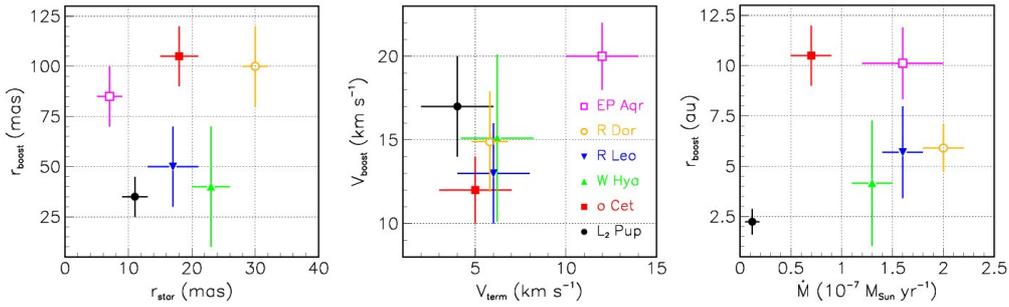

Figure 16. Location of the stars of the selected sample in the $r_{boost}$ vs $r_{star}$ (left), $V_{boost}$ vs $V_{term}$ (centre) and $r_{boost}$ vs $\dot{M}$ (right).

Ideally, one would like to tell apart, in correlation with the oscillations of the light curve, regions of the stellar disc over which gas is in-falling from regions where it is out-flowing. Such detailed information is still lacking, with the exception of W Hya (Vlemmings et al. 2017) where it is, however, only partial. Available Doppler velocity spectra measured over the stellar disc lack a systematic study of their dependence over the stellar phase and the ratio between the diameters of the beam and of the stellar disc are often too large.

Nearly fifty years have elapsed since Schwarzschild (1975) estimated the size of convective cells in red giants and mentioned the possibility that they might play a role in generating mass-loss. Later, Templeton & Karovska (2009a,b) and Kiss, Szabó, & Bedding (2006) presented indirect evidence for granulation of the photosphere from Fourier analyses of the light curves. But it was only five years ago that Paladini et al. (2018), using PIONIER on the VLT, presented clear evidence for the presence of large granulation cells on the surface of π1 Gruis. Such evidence has also been obtained, using PIONIER and NARVAL, for the red super-giant Betelgeuse (Montargès et al. 2016, López Ariste et al. 2018) and, using the latter instrument, for $o$ Ceti and χ Cygni (Fabas et al. 2011, López Ariste et al. 2019). While it is still lacking for the six stars of the present sample, ALMA observations of the continuum emission and VLT observations of the stellar disc and of its close surrounding have produced images of very high angular resolution, which support the validity of the general picture outlined by the Standard Model. Yet, clear evidence for strong inhomogeneity has been obtained for only three stars of the sample, W Hya, $o$ Cet and R Dor, and its relation with possible convective cell granulation is not clear.

The study by Vlemmings et al. (2017) of the stellar disc of W Hya and its immediate neighbourhood offers a remarkable example of inhomogeneity. The observation of continuum emission gives evidence for a small (~2×3 mas$^2$) and hot (~53,000 K) spot, correlated with a significant excess of in-falling gas seen from its CO($v$=1,3-2) emission. Two years later, observations with a broader beam (48×40 mas$^2$, Vlemmings et al. 2019a), while still giving evidence for strong inhomogeneity, did not see trace of the hot spot. Ohnaka et al. (2016,2017) using SPHERE/ZIMPOL observations at two epochs, have resolved three clumpy dust clouds at a projected distance of ~1.4-2 stellar radii, displaying clear time variations (in particular formation of a new dust cloud and disappearance of an earlier one) and correlated with the recent mass ejection described by Hoai et al. (2022b).

The case of *o* Cet is not particularly convincing when considering only VLTI/AMBER observations (Wittkovski et al. 2016) and ALMA observations (Vlemmings et al. 2019a) of continuum emission using a beam of 33×23 mas$^2$. However, Kaminski et al. (2016), using ALMA observations of the emission of the AlO(6-5) line with a similar beam (33×24 mas$^2$), have given evidence for the presence of two strong blobs of emission, at projected distances of 2.9 and 4.4 stellar radii in the NNW and E directions, respectively, revealing in-falling AlO gas having inward velocity of 4 to 7 km s$^{-1}$. However, these blobs show no obvious correlation with the outflows displayed in Figure 2.

In the case of R Dor, there is evidence for clumpy emission, both from continuum with a beam of 25×18 mas$^2$ (Vlemmings et al. 2019a) and from dust (Khouri et al. 2016a). The continuum emission of L$_2$ Pup, observed by Kervella et al. (2016) with a beam of 18×15 mas$^2$, displays some elongation along the dust disc with a western excess that the authors interpret as evidence for a planetary companion at 2.0±0.2 au; but it shows no feature that might suggest the presence of granulation. SPHERE/ZIMPOL ~650 nm observations of the disc of EP Aqr (Homan et al. 2020a) show evidence for light scattered by an inclined dust disc, but the observed intensity fluctuations are too small to claim the presence of granulation. The continuum emission of R Leo, observed by Vlemmings et al. (2019a) with a beam of 19×19 mas$^2$, reveals the outward motion of a shock wave, but no sign of hot spots; other observations (VLTI/AMBER, Wittkovski et al. 2016; VLTI/MIDI, Paladini et al. 2017; VLTI/VINCI, Fedele et al. 2005) suggest the presence of inhomogeneity and variability, but without firm evidence.

In summary, while the general features displayed by ALMA continuum and VLT optical/infrared observations unanimously support a model where the presence of large convection cells play an essential role in the generation of shocks responsible for the observed effective line broadening, direct evidence for the existence of such convective cells is scarce. When inhomogeneity is observed, usually with important variability, in the form of hot spots of emission on the stellar disc or of clumpy dust emission, relating it to the presence of convective cells remains speculative and a detailed understanding of such a relation is still lacking. New VLT and ALMA observations of high sensitivity and angular resolution are still required, in particular in relation with the stellar phase, to progress toward a more quantitative description.

*6.3 On the nature of the gas outflows*
An observation that suffers no exception is of the presence of a reservoir of molecular gas gravitationally bound to the star within ~2 stellar radii from its centre. As remarked by Vlemmings et al. (2017) for W Hya, its total mass is over three orders of magnitude larger than the mass lost by the star in the course of one pulsation period. Thus, the gas spends at least $10^3$ years in this region or even closer to the photosphere, in a shock-heated environment that significantly affects the chemistry at stake. In no case does gas escape from this reservoir in a spherical flow; it does it instead in the form of outflows covering typical solid angles at steradian scale and displaying radial distributions showing that they last several decades. An interpretation in terms of the Standard Model would suggest associating each such outflow with a giant convective cell having a lifetime of several decades, which contrasts with the millimetre continuum and VLT observations of variability on a much shorter time scale. A similar concurrence of two different time scales was observed on Betelgeuse using spectro-polarimeter observations (López Ariste et al. 2018). In some cases, as for W Hya and R Leo, the global morphology of the CSE displays modest anisotropy, suggesting that on a time scale of centuries the outflows might be emitted over 4π and anisotropy might somewhat average out. This may also be the case of EP Aqr, for which the interaction of the wind with the equatorial disc dominates and prevents a reliable evaluation. In the cases of the higher mass-loss rate, R Dor, and of the lower mass-loss rates, *o* Cet and L$_2$ Pup, the global CSE seems to retain significant anisotropy, possibly related to the orbit of Mira B in the case of *o* Cet. In such cases, the mechanism that causes the wind to retain a same anisotropy over decades or even centuries may also have its origin in the modes of pulsation and/or in the activity of convective cells.

Smaller and more episodic outflows, as observed in *o* Cet, are more prone to dissociation by interstellar UV radiation, causing SiO abundance, in contrast to CO abundance, to be confined near the star. Nhung et al. (2022) have discussed this issue in some detail but presently available observations do not allow for a reliable conclusion to be reached.

Other outflows seem to be of a different nature, looking more like mass ejections occurring over a short period of time. An extreme case is the mass ejection, which followed the magnetic flare that produced a soft X ray burst in *o* Cet in 2003 (Karovska et al. 2005); this seems, however, to be rather the exception. It is not clear whether the mass ejections observed on W Hya (left panel of Figure 5) and on R Dor (right panel of Figure 4), which look like having occurred recently over a short period of time, are of a same or different nature as the former outflows; in particular, it is not clear whether they should be described in the framework of the same hydro-dynamical models. More generally, these remarks raise the question of a possible continuity, as opposed to a completely different nature, between the mass ejection observed in *o* Cet in 2003 and the outflows observed in the other stars of the sample. Granulation, magnetic flares and coronal mass ejections are related phenomena and, in the case of the Sun, the details of these relations are still raising unanswered questions (Vasil et al. 2021).

### *6.4 Magnetic fields*

The generally accepted picture described by the Standard Model offers a framework in which to study the role played by magnetic fields in generating the nascent wind and, later, shaping it toward the planetary nebula phase. While little is known about such a possible role (Vlemmings 2019b and references therein), the presence of magnetic fields near the stellar surface has been suggested by events such as UV emission from a majority of AGB stars (Montez et al. 2017), the occurrence of a soft X-ray outburst on *o* Cet (Karovska et al. 2005) or the ephemeral presence of a small and very hot spot on the surface of W Hya (Vlemmings et al. 2017). An abundant literature addresses topics of direct relevance to the role possibly played by magnetic fields in relation to convections, pulsations and shock waves (Blackman et al. 2001, Pascoli & Lahoche 2010, Fabas et al. 2011,Thirumalai & Heyl 2013, Lèbre et al. 2014, 2015, López-Ariste et al. 2019), illustrating the importance of devoting to it increased observational attention.

Of direct relevance to the present sample of selected stars are measurements and observations made on *o* Cet, W Hya and R Leo. Significant polarization of maser emission from vibrationally excited SiO lines (Cotton et al. 2004, 2006, 2008, Herpin et al. 2006) has been measured using VLBA observations, complemented by VLTI/MIDI interferometry at 10 μm wavelength, and 30 m IRAM single dish observations. While giving clear evidence for the presence of magnetic fields of a few Gauss, they are difficult to interpret in terms of their structure, radial or toroidal. Recently, using CARMA observations, Huang et al. (2020) have measured linear polarizations from the Goldreich-Kylafis effect on the CO(2-1) emission of R Leo at the level of ~9.7%. The effect reveals unequal populations of the sublevels of the final state of the transition and gives reliable evidence for the presence of magnetic field. Finally, using NARVAL mounted on the 2 m telescope at Pic du Midi Observatory, Fabas et al. (2011) have observed variability and structures on the polarized Balmer line emission of *o* Cet, which they interpret in terms of shock waves suggesting the presence of magnetic field. Also worth mentioning is a magneto-hydrodynamic model of *o* Cet proposed by Thirumalai & Heyl (2013), which describes how an equatorial magnetic field could contribute to the production of a dust-driven wind.

### *6.5 Conclusion*

We have presented a brief review of the current knowledge of the properties displayed by a representative sample of the CSEs of nearby oxygen-rich AGB stars. Their diversity, each star appearing as a singular case, makes it difficult to identify common features with confidence. Moreover, the intrinsic limitations inherent to astronomical observations, with only two position coordinates and a single velocity component being accessible to measurement, must be kept in mind: experience shows that when more accurate measurements become available, some of our earlier preconceptions may need to be abandoned, or at least modified. We have shown that the presence of rotation near the photosphere, and of planetary or stellar companions orbiting the mass-losing star, seem to have little impact on the formation of the nascent wind, but the validity of this statement needs to be confirmed by new high angular resolution observations of stars such as R Dor, $L_2$ Pup or EP Aqr. In general, observations support current ideas on the mechanisms presiding over the formation of the nascent wind, but quantitative confirmation is still lacking. In particular the dependence on stellar phase of the morpho-kinematics at stake near the photosphere needs to be studied and the concurrence of its high variability with the persistence of anisotropy at large distances over many decades needs to be understood. The complexity of the dynamics at stake cannot be ignored and the importance of the time dependence of the mass-loss rate needs to be taken in proper account when attempting a global description of the evolution of the CSE. The outstanding performance of the VLT and of ALMA in terms of angular resolution and sensitivity will still allow for major progress over many years to come.

**Acknowledgements**
We thank Dr Bernd Freytag for clarifications concerning the Standard Model and for sharing with us his understanding of the issues developed in the present article. This paper makes use, in particular, of many archival ALMA data. ALMA is a partnership of ESO (representing its member states), NSF (USA) and NINS (Japan), together with NRC (Canada), MOST and ASIAA (Taiwan), and KASI (Republic of Korea), in cooperation with the Republic of Chile. The Joint ALMA Observatory is operated by ESO, AUI/NRAO and NAOJ. We are deeply indebted to the ALMA partnership, whose open access policy means invaluable support and encouragement for Vietnamese astrophysics. Financial support from the World Laboratory, the Vingroup Fellowship Programme, the Odon Vallet Foundation and the Vietnam National Space Center is gratefully acknowledged.

**References**
Bedding, T. R., Zijlstra, A.A., Jones, A. et al., 1998, MNRAS, 301, 1073
Bedding, T.R., Zijlstra, A.A., Jones, A. et al., 2002, MNRAS, 337, 79
Bedding, T.R., Kiss, L.L., Kjeldsen, H. et al., 2005, MNRAS, 361, 1375
Blackman, E.G., Frank, A., Markiel, J.A. et al., 2001, Nature, 409, 485
Bujarrabal, V., Alcolea, J., Mikołajewska, J. et al., 2018, A&A, 616, L3
Bujarrabal, V., Agúndez, M., Gómez-Garrido et al., 2021, A&A, 651, A4
Chandler, A.A., Tatebe, K., Wishnow, E.H. et al., 2007, ApJ, 670, 1347
Cotton, W.D., Mennesson, B., Diamond, P.J., et al., 2004, A&A, 414, 275
Cotton, W.D., Vlemmings, W.H.T., Mennesson, B. et al., 2006, A&A, 456, 339
Cotton, W.D., Perrin, G. & López, B., 2008, A&A, 477, 853
Cotton, W.D., Ragland, S., Pluzhnik, E. et al., 2009, ApJ, 704, 170
Danilovich, T., Teyssier, D., Justtanont, K. et al., 2015, A&A, 581, A60
Danilovich, T., De Beck, E., Black, J.H. et al., 2016, A&A, 588, A119
Danilovich, T., Van de Sande, M., De Beck, E. et al., 2017a, A&A, 606, A124
Danilovich, T., Lombaert, R., Decin, L. et al., 2017b, A&A, 602, A14
De Beck, E. & Olofsson, H., 2018, A&A, 6715, A8
Decin, L., Richards, A.M.S., Waters, L.B.F.M. et al., 2017, A&A, 608, A55
Decin, L., Richards, A.M.S., Danilovich, T. et al., 2018, A&A, 615, A28
Desmurs, J.F., Bujarrabal, V., Lindqvist, M. et al., 2014, A&A, 565, A127
Dumm, T. & Schild, H., 1998, New A., 3, 137
Eggen, O.J., 1973, Mem. R. Astr. Soc., 77, 159
Fabas, N., Lèbre, A. & Gillet, D., 2011, A&A, 535, A12
Fedele, D., Wittkowski, M., Paresce, F. et al., 2005, A&A, 431, 1019
Fonfría, J.P., Santander-García, M., Cernicharo, J. et al., 2019a, A&A, 622, L14
Fonfría, J.P., 2019b, poster presented at the ALMA2019: Science results and cross-facility synergies, Cagliari, October 14-18 2019, doi:10.5281/zenodo.3585368
Freytag, , Liljegren, S. and Höfner, S., 2017, A&A, 600, A137
Freytag, B. & Höfner, S., 2023, A&A, 669, A155
Gaia Collaboration, 2018, VizieR Online Data Catalog, I/345
Hadjara, M., Cruzalebes P., Nitschelm C., et al., 2019, MNRAS, 489, 2595
Herpin, F., Baudry, A., Thum, C. et al., 2006, A&A, 450, 667
Hoai, D.T., Nhung, P.T., Tuan-Anh, P. et al., 2019, MNRAS, 484, 1865
Hoai, D.T., Nhung, P.T., Tuan-Anh, P. et al., 2020a, Com. Phys. Vietnam, 30, 85
Hoai, D.T., Tuan-Anh, P., Nhung, P.T. et al., 2020b, MNRAS, 495, 943
Hoai, D.T., Tuan-Anh, P., Nhung, P.T. et al., 2021, JKAS, 54, 171

Hoai, D.T., Nhung, P.T., Darriulat, P. et al., 2022a, MNRAS, 510, 2363
Hoai, D.T., Nhung, P.T., Darriulat, P. et al., 2022b, VJSTE, 64, 16
Hoai, D.T., Nhung, P.T., Tan, M.N. et al., 2023, MNRAS, 518, 2034
Hofmann, K.-H, Balega, Y., Scholz, M. et al., 2001, A&A, 376, 518
Höfner, S. & Olofsson, H., 2018, A&Arv, 26, 1
Höfner, S. & Freytag, B., 2019, A&A, 623, A158
Homan, W., Richards, A., Decin, L. et al., 2017, A&A, 601, A5
Homan, W., Danilovich, T., Decin, L. et al., 2018a, A&A, 614, A113
Homan, W., Richards, A., Decin, L. et al., 2018b, A&A, 616, A34
Homan, W., Cannon, E., Montargès, M. et al., 2020a, A&A, 642, A93
Homan, W., Montargès, M., Pimpanuwat, B. et al., 2020b, A&A, 644, A61
Huang, K.Y., Kemball, A.J., Vlemmings, W.H.T. et al., 2020, ApJ, 899, 152
Kamiński, T., Wong, K.T., Schmidt, M.R. et al., 2016, A&A, 592, A42
Kamiński, T., Müller, H.S.P., Schmidt, M.R. et al., 2017, A&A, 599, A59
Karovska, M., Schlegel, E., Warren, H. et al., 2005, ApJ, 623, 137
Kervella, P., Montargès, M., Ridgway, S.T. et al., 2014, A&A, 564, A88
Kervella, P., Montargès, M., Lagadec, E. et al., 2015, A&A, 578, A77
Kervella, P., Homan, W., Richards, A.M.S. et al., 2016, A&A, 596, A92
Khouri, T., de Koter, A., Decin, L., et al., 2014a, A&A, 561, A5
Khouri, T., de Koter, A., Decin, L., et al., 2014b, A&A, 570, A67
Khouri, T., Waters, L.B.F.M., de Koter, A. et al. 2015, A&A, 577, A114
Khouri, T., Maercker, M., Waters, L.B.F.M. et al., 2016a, A&A, 591, A70
Khouri, T., Vlemmings, W.H.T., Ramstedt, S. et al., 2016b, MNRAS, 463, L74
Khouri, T., Vlemmings, W.H.T., Olofsson, H. et al., 2018, A&A, 620, A75
Khouri, T., Velilla-Prieto, L., De Beck, E. et al., 2019, A&A, 623, L1
Khouri, T., Vlemmings, W.H.T., Paladini, C. et al., 2020, A&A, 635, A200
Kipper, T., 1992, Baltic Astronomy, 1, 190
Kiss, L.L., Szabó, G., & Bedding, T.R., 2006, MNRAS, 372, 1721
Knapp, G. R., Pourbaix, D., Platais, I. et al., 2003, A&A, 403, 993
Lèbre, A., Aurière, M., Fabas, N. et al., 2014, A&A, 561, A85
Lèbre, A., Aurière, M., Fabas, N. et al., 2015, Proc. of the IAU, 305, 47
Lebzelter, T. & Hron, J., 1999, A&A, 351, 533
Liljegren, S., Höfner, S., Freytag, B. & Bradh, S., 2018, A&A, 619, A47
Little, S.J., Little-Marenin, I.R. & Bauer, W.H., 1987, AJ, 94, 981
López Ariste, A., Mathias, P., Tessore, B. et al., 2018, A&A, 620, A199
López-Ariste, A., Tessore, B., Carlin, E.S. et al., 2019, A&A, 632, A30
Lykou, F., Klotz, D., Paladini, C. et al., 2015, A&A, 576, A46, erratum A&A, 581, C2
Matthews, L.D., Reid, M.J. & Menten, K.M., 2015, ApJ, 808, 36
Matthews, L.D., Reid, M.J., Menten, K.M. et al., 2018, AJ, 156, 15
McIntosh, G.C. & Indermühle, B., 2013, ApJ, 774, 21
Montargès, M., Kervella, P., Perrin, G., et al., 2016, A&A, 588, A130
Montez, R.Jr., Ramstedt, S., Kastner, J.H., Vlemmings, W., & Sanchez, E., 2017, ApJ, 841, 33
Morris, M., Raghvendra, S., Matthews, K. et al., 2006, Proceedings of IAU, 234, 469. Edited by M.J. Barlow & R.H. Méndez, Cambridge University Press.
Nhung, P.T., Hoai, D.T., Diep, P.N. et al., 2016, MNRAS, 460, 673
Nhung, P.T., Hoai, D.T., Tuan-Anh, P. et al., 2019, MNRAS, 490, 3329
Nhung, P.T., Hoai, D.T., Tuan-Anh, P. et al., 2021, MNRAS, 504, 2687
Nhung, P.T., Hoai, D.T., Tuan-Anh, P. et al., 2022, ApJ, 927,169
Nhung, P.T., Hoai, D.T., Darriulat, P. et al., 2023a, RAA, 23, 015004
Nhung, P.T., Hoai, D.T., Darriulat, P. et al., 2023b, submitted to MNRAS
Norris, B.R.M., Tuthill, P.G., Ireland, M.J. et al., 2012, Nature, 484, 220
Ohnaka, K., Schertl, D., Hofmann, K.H. & Weigelt, G., 2015, A&A, 581, A127
Ohnaka, K., Weigelt, G. & Hofmann, K.H., 2016, A&A, 589, A91
Ohnaka, K., Weigelt, G. & Hofmann, K.H., 2017, A&A, 597, A20
Ohnaka, K., Weigelt, G. & Hofmann, K.H., 2019, ApJ, 883, 89
Paladini, C., Klotz, D., Sacuto, S. et al., 2017, A&A, 600, A136
Paladini, C., Baron, F., Jorissen, A. et al., 2018, Nature, 553, 310
Pascoli, G. & Lahoche, L., 2010, PASP, 122, 1334
Perrin, G., Coudé du Foresto, V., Ridgway, S.T. et al., 1999, A&A, 345, 221

Perrin, G., Ridgway, S.T., Lacour, S. et al., 2020, A&A, 642, A82
Planesas, P., Alcolea, J. & Bachiller R., 2016, A&A, 586, A69
Pojmanski, G., Maciejewski, G., Pilecki, B. et al., 2005, VizieR On-line Data Catalog: II/264
Privitera, G., Meynet, G., Eggenberger, P. et al., 2016, A&A, 593, 128
Ramstedt, S., Mohamed, S., Vlemmings, W.H.T. et al., 2014, A&A, 570, L14
Randall, S.K., Trejo, A., Humphreys, E.M.L. et al., 2020, A&A, 636, A123
Reid, M.J. & Menten, K.M, 2007, ApJ, 671, 2068
Rizzo, J.R., Cernicharo, J. & Garciá-Míro, C., 2021, ApJS, 253, 44
Ryde, N., Schoeier, F.L., 2001, ApJ, 547, 384
Samus, N.N, 2004, VizieR On-line Data Catalog: II/250.
Samus, N.N, 2009, VizieR On-line Data Catalog: B/GCVS.
Schmid, H.M., Bazzon, A., Milli, J. et al., 2017, A&A, 602, A53
Schwarzschild, M., 1975, ApJ, 195, 137
Soria-Ruiz, R., Alcolea, J., Colomer, F. et al., 2007, A&A, 468, L1
Tabur, V., Bedding, T.R., Kiss, L.L. et al., 2009, MNRAS, 400, 1945
Takigawa, A., Kamizuka, T., Tachibana, S. et al., 2017, Science Advances, 3/11, eaao2149
Takigawa, A., Kim, T.H., Igami, Y. et al., 2019, ApJ, 878, L7
Tatebe, K., Wishnow, E.H., Ryan, C.S. et al., 2008, ApJ, 689, 1289
Templeton, M.R. & Karovska, M., 2009a, AIP Conference Proceedings, 1170, 164
Templeton, M.R. & Karovska, M., 2009b, ApJ, 691, 1470
Thirumalai, A. & Heyl, J.S., 2013, MNRAS, 430, 1359
Tuan-Anh, P., Hoai, D.T., Nhung, P.T. et al., 2019, MNRAS, 487, 622
Van Langevelde H., et al., 2019, Proceeding of 14th European VLBI Network Symposium & Users Meeting (EVN 2018), 344, 43 (arXiv:1901.07804)
Van Leeuwen, F., 2007, *Hipparcos, the New Reduction of the RawData*, Springer, Astrophysics and Space Science Library, vol.350
Vanture, A.D., Wallerstein, G., Brown, J.A. & Bazan, G., 1991, Bulletin of the American Astronomical Society, 23, 966
Vasil, G.M., Julien, K. & Featherstone, N.A. , 2021, Proceedings of the National Academy of Sciences, 118, 31, doi: 10.1073/pnas.2022518118
Vlemmings, W.H.T., Ramstedt, S., O'Gorman E. et al., 2015, A&A, 577, L4
Vlemmings, W.H.T., Khouri, T., O'Gorman, E. et al., 2017, Nature Astronomy, 1, 848
Vlemmings, W.H.T., Khouri, T., De Beck, E. et al., 2018, A&A, 613, L4
Vlemmings, W.H.T., Khouri, T. & Olofsson, H., 2019a, A&A, 626, A81
Vlemmings, W.H.T., 2019b, Proceedings of IAU, 343, 19
Vlemmings, W.H.T., Khouri, T. & Tafoya, D., 2021, A&A, 654, A18
Watson, C.L., 2006, JAAVSO, 35, 318
Wenger, M., Ochsenbein, F., Egret, D. et al., 2000, A&AS, 143, 9
Wiesemeyer, H.W., Thun, C., Baudry, A. & Herpin, F., 2009, A&A, 498, 3
Winters, J.M., Le Bertre, T. & Nyman, L.A., 2002, A&A, 388, 609
Wittkowski, M., Chiavassa, A., Freytag, B. et al., 2016, A&A, 587, A12
Wong, K.T., Kamínski, T., Menten, K.M. & Wyrowski, F., 2016, A&A, 590
Woodruff, H. C., Ireland, M. J., Tuthill, P. G. et al., 2009, ApJ, 691, 1328
Woodruff, H. C., Tuthill, P. G., Monnier, J. D. et al., 2008, ApJ, 673, 418
Zhao-Geisler, R., Quirrenbach, A., Köhler, R. et al., 2012, A&A, 545, A56
Zhao-Geisler, R., Koehler, R., Kemper F., et al., 2015, PASP, 127, 732

# Appendix

Table A1. Parameters and references of relevance to the six selected stars. References are indicated as the first four letters of the name of the first author, followed by the last two digits of the year of publication. Spectral and variability types are taken from standard catalogues (*Weng00, Samu04, Samu09*). Distance (*d*) measurements are typically reliable to within ±10%, with the exception of R Leo for which we retain the Hipparcos value of 114 pc (*vanL07*) rather than the Gaia value of 71 pc (*Gaia18*) (see van Langevelde et al. 2019). Stellar radii ($R_{star}$) measurements span broad samples of values, depending on stellar phase and wavelength. We retain values (continuum at millimetre wavelengths and near infrared) quoted by *Vlem19a* in a reliable and critical review of four of the stars in our sample. For the other two stars ($L_2$ Pup and EP Aqr), the continuum values are poorly measured. The period of EP Aqr is poorly known and often quoted as 55 days while it is in fact more like 110 days. Masses are not precisely measured and are mostly crude estimates.

|  | *$L_2$ Pup* *HD56096* | *o Cet* *HD14386* | *W Hya* *HD120285* | *R Leo* *HD84748* | *R Dor* *HD29712* | *EP Aqr* *HD207076* |
|---|---|---|---|---|---|---|
| *Type (spectral /variable)* | M5IIIe /SRb | M5-M&IIIe /Mira | M7.5-9 /SRa | M7-9 /Mira | M8 /SRb | M7-8 /SRb |
| *d (pc)* | 64±4 VanL07 | ~100 VanL07 | 104 VanL07 | 114 VanL07 | 59 Knap03 | 119 VanL07 Gaia18 |
| *$R_{star}$ (mas)* continuum/NIR | ?/9-10 Kerv14 Ohna15 | 21/15 Vlem19a | 26/20 Vlem19a | 21/14 Vlem19a | 32/28 Vlem19a | ~8 /6 Homa20a Dumm98 |
| *Period (d)* | 141 Bedd02,05 | 333 Temp09a,b | 361 Samu04 | 310 Wats06 | 175/332 Bedd98 | 110 Egge73, Tabu09 |
| *Mass ($M_{Sun}$)* | ~0.7 Kerv16 | A~2.0, B~0.7 Plan16 | 1.0-1.5 Khou14a,b, Dani17b | ~0.7 Wies09 | 0.7-1.0 Ohna19 | ~1.7 Homa20a |
| *Mass-loss rate ($10^{-7} M_{Sun}/yr$)* | 0.12±0.07 Hoai22a | 0.7±0.2/2.5 Nhun22/Ryde01 | 1.3 Khou14a,b | ~1.6 Dani15 | ~2 DeBe18 | 1.6±0.4 Hoai19 |
| *Terminal velocity (km s$^{-1}$)* | 3.4±1.7 Hoai22a | 2.5 Ryde01 | 5 Hoai22b | 5.5 Hoai23 | 6-9 Nhun19 | 2-11 Hoai19 |
| *VLT NACO VLTI VINCI* | Kerv14 Lyko15 Ohna15 |  | Norr12 | Norr12 Fede05 | Norr12 |  |
| *VLT SPHERE* | Kerv15 | Khou18 | Ohna16,17 Khou20 |  | Khou16a | Homa20a |
| *VLTI AMBER* | Ohna15 | Witt16 | Ohna16,17 Hadj19 | Witt16 | Ohna19 |  |
| *VLTI MIDI* | Kerv14 |  | Zhao12,15 | Pala17 |  |  |
| *Others* |  | Chan07/ISI Perr20/IOTA Tate08/ISI Wood08,09/Keck | Wood08,09/Keck Khou15/ISO | Perr99/IOTA Hofm01/SAO Wood08,09/Keck Tate08/ISI |  |  |
| *ALMA Continuum* | Kerv16 | Matt15 Vlem15,19a Plan16 Kami16 Wong16, Khou18 | Vlem17,19a | Vlem19a | Vlem19a | Homa20a |
| *ALMA and APEX Abundances* |  | Kami16,17 | Taki17 Dani17a,b Khou19 |  | Dani16 Deci17 Khou19 |  |
| *ALMA Molecular lines* | Kerv16 Homa17 Hoai22a | Hoai20b Khou16b,18 Wong16 Nhun16,22 | Taki17 Vlem17 Hoai21 Hoai22b | Fonf19a,b Hoai23 | Vlem18 Deci18 Hoai20a Homa18a Nhun19,21 | Homa18b Tuan19 Hoai19 Homa20a Nhun23b |
| *Others* |  | Matt15/JVLA |  | Reid07/VLA Matt18/JVLA | DeBe18 APEX/SEPIA |  |
| *Masers* | McIn13 | Cott04,06,08 Rizz21 | Vlem21, Cott08 Rizz21, Herp06 | Cott04,08,09 Rizz21, Sori07 Desm14, Herp06 |  |  |

Table A2. Molecular lines observed and studied with high angular resolution (beam FWHM<~200 mas) in the selected sample of six stars. Emission related parameters (Einstein coefficient $A$, energy of the upper level $E_{up}$ and frequency $f$ of the transition) are listed in the first columns. For each star the mean beam size (FWHM) associated with the highest angular resolution is listed in mas in the last columns. We recall that the emissivity $\varepsilon$ and the optical depth $\tau$ are obtained from the relations $\varepsilon[\text{Jy}\times\text{arcsec}^{-2}]/N[\text{mol cm}^{-3}\text{arcsec}]=5.5\times10^3\times d[\text{pc}]\times A[\text{s}^{-1}]\,f_{pop}$, $f_{pop}=k(2J+1)e^{-E_u/T}/T$ and $\varepsilon/\tau=0.34\times10^{-7}f^3/(e^{\Delta E/T}-1)$ where $N$ is the density, $d$ the distance from Earth to the star and $k$ is the product of the partition function by the temperature.

| Line | $A$ [Hz] | $E_{up}$ [K] | $f$ [GHz] | L₂ Pup | o Cet | W Hya | R Leo | R Dor | EP Aqr |
|---|---|---|---|---|---|---|---|---|---|
| $^{12}C^{16}O(v=0,2-1)$ | $6.91\times10^{-7}$ | 16.6 | 230.5 | | | | 40 | | 25 |
| $^{13}C^{16}O(v=0,2-1)$ | $6.04\times10^{-7}$ | 15.9 | 220.4 | | | | | | 23 |
| $^{12}C^{16}O(v=0,3-2)$ | $2.50\times10^{-6}$ | 33.2 | 345.8 | 21 | | 46 | | 160 | |
| $^{12}C^{16}O(v=1,3-2)$ | $1.45\times10^{-6}$ | 3120 | 342.6 | | | 17 | | | |
| $^{13}C^{16}O(v=0,3-2)$ | $2.18\times10^{-6}$ | 31.7 | 330.6 | 22 | | | | | |
| $^{28}Si^{16}O(v=0,5-4)$ | $5.91\times10^{-4}$ | 31.3 | 217.1 | | 32-45 | | | | 24 |
| $^{28}Si^{16}O(v=1,5-4)$ | $0.50\times10^{-3}$ | 1800 | 215.6 | | 32 | | 42 | | |
| $^{28}Si^{16}O(v=1,8-7)$ | $2.20\times10^{-3}$ | 1844 | 344.9 | | | | 47 | 150 | |
| $^{28}Si^{16}O(v=2,5-4)$ | $0.51\times10^{-3}$ | 3552 | 214.1 | | 32 | | 46 | | |
| $^{28}Si^{16}O(v=3,5-4)$ | $5.08\times10^{-4}$ | 5287 | 212.6 | | | | | 36 | |
| $^{29}Si^{16}O(v=0,5-4)$ | $0.50\times10^{-3}$ | 30.9 | 214.4 | | 32 | | 46 | | |
| $^{28}Si^{16}O(v=0,8-7)$ | $2.20\times10^{-3}$ | 75.0 | 347.3 | | | | | 30 | |
| $^{29}Si^{16}O(v=1,5-4)$ | $4.93\times10^{-4}$ | 1789 | 212.9 | | | | | 36 | |
| $^{29}Si^{16}O(v=0,8-7)$ | $2.10\times10^{-3}$ | 74.1 | 343.0 | 20 | | 35-45 | | 30 | |
| $^{30}SiO(v=0,8-7)$ | $2.05\times10^{-3}$ | 73.2 | 338.9 | | | | | 150 | |
| $^{28}Si^{32}S(v=1,19-18)$ | $6.89\times10^{-4}$ | 1236 | 343.1 | 20 | | | | | |
| $^{32}S^{16}O(5_5,4_4)$ | $0.12\times10^{-3}$ | 44.1 | 215.2 | | | | 42 | | |
| $^{32}S^{16}O(6_5,5_4)$ | $1.96\times10^{-4}$ | 50.7 | 251.8 | | | | | 150 | |
| $^{32}S^{16}O(v=0,8_8-7_7)$ | $5.19\times10^{-4}$ | 87.5 | 344.3 | 20 | | | | | |
| $^{32}S^{16}O(v=0,8_9-7_8)$ | $5.38\times10^{-4}$ | 78.8 | 346.5 | | | | | 150 | |
| $^{32}S^{16}O_2(34_{3,31},34_{2,32})$ | $3.45\times10^{-4}$ | 581.9 | 342.8 | | | | | 38 | |
| $^{32}S^{16}O_2(16_{3,13},16_{2,14})$ | $9.90\times10^{-5}$ | 147.8 | 214.7 | | | | | 36 | |
| $^{32}S^{16}O_2(16_{6,10},17_{5,13})$ | $2.35\times10^{-5}$ | 213.3 | 234.4 | | | | | | 175 |
| $^{32}S^{16}O_2(13_{2,12},12_{1,11})$ | $2.38\times10^{-4}$ | 92.8 | 345.3 | | | | 47 | | |
| $^{32}S^{16}O_2(22_{2,20},22_{1,21})$ | $9.30\times10^{-5}$ | 248.4 | 216.6 | | | | 42 | | |
| $^{32}S^{16}O_2(4_{2,2}-3_{1,3})$ | $7.69\times10^{-5}$ | 19.0 | 235.2 | | | | | | 25 |
| $H_2^{16}O(v_2=1,5_{5,0},6_{4,3})$ | $4.77\times10^{-6}$ | 3462 | 232.7 | | 30 | | | | |
| $H^{12}C^{14}N(4-3)$ | $2.06\times10^{-3}$ | 42.5 | 354.5 | | | | | 151 | |
| $H^{13}C^{14}N(4-3)$ | $1.90\times10^{-3}$ | 41.4 | 344.2 | | | | 45 | | |
| $^{27}Al^{16}O(N=9,8)$ | See Taki17 | | 344.4 | | | 34 | | | |
| $^{16}OH$ Many (see Khou19) | | | | | | ~30 | | ~30 | |